\documentclass[twocolumn,ngerman,english,prb, showpacs]{revtex4-1}
\usepackage[T1]{fontenc}
\usepackage[latin9]{inputenc}
\usepackage{amsmath}
\usepackage{amssymb}
\usepackage{graphicx}

\makeatletter

\newcommand*\LyXThinSpace{\,\hspace{0pt}}

\makeatother

\usepackage{babel}
\begin{document}

\title{\global\long\def\sh{\textrm{sh}}
\global\long\def\ch{\textrm{ch}}
\global\long\def\th{\textrm{th}}
Many-body theory of magneto-elasticity in one dimension}

\author{O. Tsyplyatyev}
\selectlanguage{ngerman}%

\affiliation{Institut für Theoretische Physik, Goethe-Universität Frankfurt, Max-von-Laue
Strasse 1, 60438 Frankfurt am Main, Germany}
\selectlanguage{english}%

\author{P. Kopietz}
\selectlanguage{ngerman}%

\affiliation{Institut für Theoretische Physik,  Goethe-Universität Frankfurt, Max-von-Laue
Strasse 1, 60438 Frankfurt am Main, Germany}

\author{Y. Tsui}

\affiliation{Physikalisches Institut, Goethe-Universität Frankfurt, Max-von-Laue-Strasse
	1, 60438 Frankfurt am Main, Germany}

\author{B. Wolf}

\affiliation{Physikalisches Institut, Goethe-Universität Frankfurt, Max-von-Laue-Strasse
	1, 60438 Frankfurt am Main, Germany}

\author{P. T. Cong}

\altaffiliation[present address: ]{Dresden High Magnetic Field Laboratory, Helmholtz-Zentrum Dresden-Rossendorf, D-01314 Dresden, Germany.}

\author{N. van Well}


\altaffiliation[present address: ]{Laboratory for Neutron Scattering and Imaging, Paul Scherrer Institute, 5232 Villigen, Switzerland.}

\author{F. Ritter}

\affiliation{Physikalisches Institut, Goethe-Universität Frankfurt, Max-von-Laue-Strasse
	1, 60438 Frankfurt am Main, Germany}

\author{C. Krellner}

\affiliation{Physikalisches Institut, Goethe-Universität Frankfurt, Max-von-Laue-Strasse
	1, 60438 Frankfurt am Main, Germany}

\author{W. A\ss mus}

\affiliation{Physikalisches Institut, Goethe-Universität Frankfurt, Max-von-Laue-Strasse
	1, 60438 Frankfurt am Main, Germany}

\author{M. Lang}

\affiliation{Physikalisches Institut, Goethe-Universität Frankfurt, Max-von-Laue-Strasse
	1, 60438 Frankfurt am Main, Germany}

\date{\today}
\selectlanguage{english}%
\begin{abstract}
We construct a many-body theory of magneto-elasticity in one dimension
and show that the dynamical correlation functions of the quantum magnet,
connecting the spins with phonons, involve all energy scales. Accounting
for all magnetic states non-perturbatively via the exact diagonalisation
techniques of Bethe ansatz, we find that the renormalisation of the
phonon velocity is a non-monotonous function of the external magnetic
field and identify a new mechanism for attenuation of phonons \textendash{}
via hybridisation with the continuum of excitations at high energy.
We conduct ultrasonic measurements on a high-quality single crystal of the frustrated spin-1/2 Heisenberg antiferromagnet $\textrm{Cs}_{2}\textrm{CuCl}_{4}$
in its nearly one-dimensional regime and confirm the theoretical predictions,
demonstrating that ultrasound can be used as a powerful probe of strong correlations in one dimension.
\end{abstract}


\maketitle
\section{Introduction}
Magnetic insulators present a good example of interacting
quantum systems where phonons can serve as an intrinsic probe of the strongly-correlated
spins. \cite{Luethi_book} The first microscopic theory of magneto-elasticity
was developed at finite temperatures,\cite{Kawasaki70,Tani68} where
the  static and the dynamic correlation functions of the spins
were shown to couple to phonons with the same strength in the perturbative regime.
At low temperature, assuming existence of a spin-liquid regime in two- and three-dimensional Heisenberg antiferromagnets, phonons were shown to measure the mass and lifetime of the spin-liquid quasiparticles.
\cite{Zhou11,Serbyn13} In one dimension (1D) \textendash{} where
interacting magnons form a spin-Luttinger liquid at low energy\cite{Giamarchi_book}
\textendash{} the theory remains largely unexplored. At the same time
such 1D systems are readily accessible in experiments on $\textrm{Cs}_{2}\textrm{CuCl}_{4}$,
\cite{Coldea03,Sytcheva09} $\textrm{CsNiCl}_{3}$,\cite{Trudeau92} $\textrm{KCuF}_{3}$,\cite{Bella05} and a metal organic coordination polymer Cu(II)-2,5-bis(pyrazol-1-yl)-1,4-dihydroxybenzene. \cite{Wolf04}

In this paper we construct a microscopic theory of magneto-elasticity
in 1D using the diagonalisation methods of Bethe ansatz.\cite{Gaudin_book}
We derive the matrix elements for the four-point correlation function
that couples the strongly-correlated spins to phonons dynamically
and show that Luttinger liquid at low energy contributes comparably
with the high-energy excitations that we are able to account for due
to the hierarchy of modes.\cite{OT15,OT16,Moreno16} The contribution of the
static correlation function to the renormalisation of the sound velocity
is parametrically larger than the dynamical correlation functions.
The resonant decay of phonons in the many-body spin continuum vanishes
very fast, as the fourth power of the length in large systems. However
we identify another mechanism, hybridisation with the excitations
at high energy via the dynamical correlation function, that remains finite in the thermodynamic limit. This
work advances the many-body diagonalisation tools in 1D\cite{Kitanine00,Kitanine99,JES05,JES07,Bella13}
in to the field of magneto-elasticity, which is beginning  to receive attention also in spintronics. \cite{Bauer14,Bauer15,Kopietz14,Kopietz15_2}

To test our theory we conduct ultrasonic measurements on a high-quality
single crystal of $\textrm{Cs}_{2}\textrm{CuCl}_{4}$ in its nearly 1D regime,
\emph{i.e.} at temperatures of $0.7-2.1$K and magnetic fields up to $9$T. \cite{Coldea97,Balents07}
The observed dependencies of the sound velocity and attenuation of
the sound wave on the magnetic field agree well with all theoretical
predictions. We find that the magnetic-field dependent part of the attenuation
is governed by the hybridisation mechanism.  Our results demonstrate that ultrasonic investigations, besides neutron-scattering experiments, \cite{Coldea03,Bella05, Birgeneau72} can be used as a powerful probe of correlation functions of the many-body system in 1D in magnetic insulators, just as tunnelling spectroscopy in semiconductor heterostructures. \cite{Ford09,Yacoby02}

The paper is organised as follows. Section II contains definition of the magnetostrictive interactions between the Heisenberg model and the phonon models in one dimensions and the diagonlisation of the isolated Heisenberg model by means of Bethe ansatz. In Section III we study renomalisation of sound velocity by evaluating microscopically the dynamical correlation function of the spins that couples to the phonons (Subsection IIIA) and by analysing it using the hierarchy of interacting modes (Subsection IIIB). In Section IV we consider different mechanisms of attenuation of phonons. And in Section V we conduct an ultrasound experiment on $\textrm{Cs}_{2}\textrm{CuCl}_{4}$ in its nearly one-dimensional regime and confirm the theoretical predictions. In Appendix A we derive the quantisaion equation for the pi-pairs' solutions of Bethe equations in the XY limit. In Appendix B we quote the normalisation factor of the Bethe states  together with the algebraic Bethe ansatz method. And in Appendix C we derive the matrix element of the spin operator needed for the magnetostrictive interaction.

\section{Model}
Theoretically, we consider phonons interacting with 1/2-spins on a 1D lattice of length $L$ via
a magnetostrictive interaction as\cite{Kawasaki70,Tani68} 
\begin{equation}
H=H_{m}+H_{ph}+V,
\end{equation}
where
\begin{align}
H_{m} & =\sum_{{\scriptstyle j=1}}^{{\scriptstyle L}}(J\mathbf{S}_{{\scriptscriptstyle j}}\cdot\mathbf{S}_{{\scriptscriptstyle j+1}}+BS_{{\scriptscriptstyle j}}^{{\scriptstyle z}}),\;\,H_{ph}=\sum_{{\scriptstyle k}}\omega_{{\scriptscriptstyle k}}a_{{\scriptscriptstyle k}}^{\dagger}a_{{\scriptscriptstyle k}},\label{eq:H_m_ph}\\
V & =\sum_{{\scriptstyle j=1}}^{{\scriptstyle L}}[J_{1}(x_{{\scriptscriptstyle j+1}}-x_{{\scriptscriptstyle j}})+J_{2}(x_{{\scriptscriptstyle j+1}}-x_{{\scriptscriptstyle j}})^{2}]\mathbf{S}_{{\scriptscriptstyle j}}\cdot\mathbf{S}_{{\scriptscriptstyle j+1}},\label{eq:V}
\end{align}
are the Heisenberg model of spins, the free phonon model, and the
interaction between them, respectively, $\mathbf{S}_{j}$ are the spin-1/2
operators, $J$ is the exchange interaction between spins when the
atoms are in equilibrium, and $B$ is the external magnetic field in energy
units. Here $a_{k}$ are Bose operators of the phonons, $\omega_{k}=2\omega_{D}\left|\sin\left(k/2\right)\right|$
is the phonon dispersion, $\omega_{D}$ is Debye energy, $x_{j}=\sum_{k}\sqrt{\frac{\hbar b}{mv_{0}\left|k\right|L}}(a_{k}+a_{-k}^{\dagger})e^{-ikj}$
is the position operator of an atom with the mass $m$ at lattice site $j$, and $v_{0}$ is the sound velocity. Phononic
excitations modulate the exchange integrals resulting in a set of magnetostrictive
constants $J_{n}=\left.\partial_{x}^{n}J\left(x\right)\right|_{x=b}/n!$ that quantify the magneto-elastic
interaction,
where $b$ is the lattice parameter. We assume the periodic boundary condition: $\mathbf{S}_{j+L}=\mathbf{S}_{j}$
and $x_{j+L}=x_{j}$.

The spin Hamiltonian in Eq. (\ref{eq:H_m_ph}) is diagonalised by
$N$-magnon states parameterised with a set of $N$ quasimomenta $\mathbf{q}=(q_{1}\dots q_{N})$
that satisfy the non-linear Bethe equations \cite{Gaudin_book} 
\begin{equation}
q_{j}L-\sum_{l\neq j}\varphi_{jl}=2\pi I_{j},
\end{equation}
where the two-body scattering phases are 
\begin{equation}
e^{i\varphi_{ij}}=-\frac{e^{i\left(q_{i}+q_{j}\right)}+1-2\Delta e^{iq_{i}}}{e^{i\left(q_{i}+q_{j}\right)}+1-2\Delta e^{iq_{j}}},
\end{equation}
$\Delta=1$, and $I_{j}$ is a set of non-equal integers. Solutions
of Bethe equations can be found via numerical deformation from the XY point
$\Delta=0$ (where $\varphi_{ij}=\pi$ gives the solutions
$q_{j}=2\pi I_{j}/L$) to the Heisenberg point $\Delta=1$. \cite{strings} However, Bethe equations remain non-linear, $\alpha L-\Phi\left(\alpha,\mathbf{q}\right)=2\pi I_{j}$,
for some solutions that contain at least a pair of quasimomenta satisfying
the condition $q_{i}+q_{j}=\pm\pi$ in the $\Delta=0$ limit \textendash{}
see derivation in Appendix A. Here $q_{i}=\pm\pi-\alpha$,
$q_{j}=\alpha$, the scattering phase 
\begin{equation}
e^{i\Phi\left(\alpha,\mathbf{q}\right)}=-\frac{i\frac{2\lambda}{L}\sum_{j=1}^{N-2r}\frac{1-\sin q{}_{j}\sin\alpha}{\sin q_{j}-\sin\alpha}+e^{i\alpha}}{i\frac{2\lambda}{L}\sum_{j=1}^{N-2r}\frac{1-\sin q{}_{j}\sin\alpha}{\sin q_{j}-\sin\alpha}-e^{-i\alpha}}\label{eq:pi_pairs}
\end{equation}
depends on another quasimomenta, $n$ is the number of such pi-pairs,
and $\lambda=1$. Solutions for $\alpha$ can be obtained again via
deformation from the $\lambda=0$ to the $\lambda=1$ point. The eigenenergy
of $H_{m}$ corresponding to the state $\mathbf{q}$ is 
\begin{equation}
\varepsilon=\sum_{j=1}^{N}(J\cos q_{j}-J+B)+\left(\frac{J}{2}-B\right)\frac{L}{2}
\end{equation}
and the total momentum \textendash{} preserved by the translational
invariance \textendash{} is $Q=\sum_{j=1}^{N}q_{j}$.

We consider renormalisation of phonons by spins via the magnetostrictive
interaction $V$ perturbatively. The  perturbation series for
the eigenenergy of $H$ is 
\begin{equation}
E\left(k\right)=\varepsilon_{0}+\omega_{k}+\left\langle k|V|k\right\rangle +\sum_{\left\{ \mathbf{k},\mathbf{q}\right\} }\frac{\left|\left\langle \mathbf{k},\mathbf{q}|V|k\right\rangle \right|^{2}}{\varepsilon_{0}+\omega_{k}-\varepsilon_{\mathbf{q}}-\omega_{\mathbf{k}}},
\end{equation}
where $\varepsilon_{0}$ is the ground state energy of $H_{m}$, $\omega_{\mathbf{k}}$ is an eigenenergy of $H_{ph}$ parameterised
by $M$ momenta $\mathbf{k}=(k_{1},\cdots,k_{M})$. The unperturbed
state $\left|k\right\rangle =\left|k\right\rangle _{ph}\left|0\right\rangle _{m}$
is a direct product of a single phonon $\left|k\right\rangle _{ph}$
and the spin ground state $\left|0\right\rangle _{m}$ and $\left|\mathbf{k},\mathbf{q}\right\rangle =\left|\mathbf{k}\right\rangle _{ph}\left|\mathbf{q}\right\rangle _{m}$
are the intermediate states. 
\begin{figure}[b]
	\begin{center}\includegraphics[width=0.9\columnwidth]{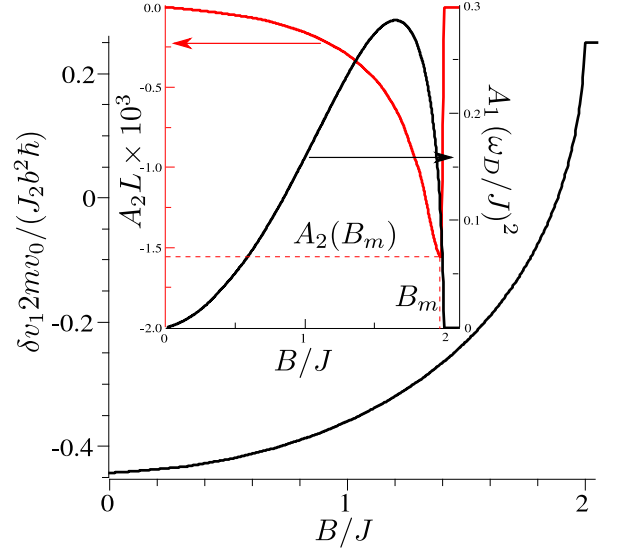}\end{center}\caption{\label{fig:static_corr}The static spin correlation function
		from Eq. (\ref{eq:dv1}) as a function of the magnetic field $B$,
		calculated using  Bethe ansatz. Inset is the dynamic
		correlation function from Eq. (\ref{eq:dv2}) calculated using the matrix element in Eq. (\ref{eq:S1S2_dynamic_matrix_element}):
		the black line is the Luttinger liquid contribution in Eq. (\ref{eq:A1})
		and the red line is the high-energy contribution in Eq. (\ref{eq:A23}). }
\end{figure}

\section{Renormalisation of sound velocity}
Change of the sound velocity is given by a  derivative of $E(k)$ as 
\begin{equation}
\delta v=\delta v_{1}+\delta v_{2},
\end{equation}
where evaluation of the phononic matrix elements leaves the spin
correlation functions split into the static and the dynamic parts,
\begin{align}
\delta v_{1} & =\frac{J_{2}b^{2}\hbar}{2mv_{0}}\left\langle 0|\mathbf{S}_{1}\cdot\mathbf{S}_{2}|0\right\rangle _{m},\label{eq:dv1}\\
\delta v_{2} & =\frac{J_{1}^{2}b^{2}\hbar}{mv_{0}}\sum_{\left\{ \mathbf{q}\right\} ;Q=p}\frac{L\left|\left\langle \mathbf{q}|\mathbf{S}_{1}\cdot\mathbf{S}_{2}|0\right\rangle _{m}\right|^{2}\left(\varepsilon_{0}-\varepsilon_{\mathbf{q}}\right)}{\left(\varepsilon_{0}-\varepsilon_{\mathbf{q}}\right)^{2}-\left(\omega_{D}p\right)^{2}}.\label{eq:dv2}
\end{align}
Here $\hbar p/b=2\pi\hbar/\left(bL\right)$ is the quantum of the
momentum and the sum over all of the many-magnon states, $\left\{ \mathbf{q}\right\} $,
is restricted by  momentum conservation to the states with $Q=p$. The static correlation
function in $\delta v_{1}$ is immediately obtained from $\varepsilon_{0}$
using the translational invariance as\cite{Orbach58} 
\begin{equation}
\left\langle 0|\mathbf{S}_{1}\cdot\mathbf{S}_{2}|0\right\rangle _{m}=\frac{\varepsilon_{0}-B\left(N-\frac{L}{2}\right)}{JL}.
\end{equation}
Its dependence on the magnetic field, changing from
the ferromagnetic value of $0.25$ at high fields to the antiferromagnetic
$\simeq-0.44$ in zero field, is shown in Fig. \ref{fig:static_corr}. 

\subsection{Dynamical correlation function of spins}
The dynamical part in $\delta v_{2}$ is a fourth-order correlation
function. We evaluate the needed matrix element using the algebraic
Bethe ansatz\cite{Korepin_book_main} and obtain it as a sum over determinants of $N\times N$
matrices, see Appendix C for details, \begin{widetext}
\begin{multline}
\left\langle \mathbf{q}|\mathbf{S}_{1}\cdot\mathbf{S}_{2}|0\right\rangle =\left(\sqrt{\left\langle 0|0\right\rangle \left\langle \mathbf{q}|\mathbf{q}\right\rangle }\right)^{-1}\Bigg\{\frac{\underset{i}{\prod}\ch\left(v_{j}+\eta\right)}{\underset{i<j}{\prod}\sh\left(v_{i}-v_{j}\right)}\sum_{y}(-1)^{y}\frac{\underset{i,j;j\neq y}{\prod}\sh\left(u_{j}-v_{i}\right)}{\underset{j}{\prod}\ch^{2}\left(u_{j}-\eta\right)}\prod_{l;l\neq y}\frac{\sh\left(u_{l}-u_{y}+2\eta\right)}{\sh\left(u_{l}-u_{y}\right)}\allowdisplaybreaks\\
\times\Bigg[\det\hat{K}^{\left(y\right)}-\Bigg(1-\frac{2\,\sh\left(2\eta\right)\sh\eta\,\sh u_{y}\underset{j;j\neq y}{\prod}\ch\left(u_{j}+\eta\right)}{\underset{i<j\neq y}{\prod}\sh\left(u_{i}-u_{j}\right)}\Bigg)\det\hat{G}^{\left(y\right)}\Bigg]-\frac{\underset{j}{\prod}\ch\left(u_{j}+\eta\right)\underset{j}{\prod}\ch\left(v_{j}+\eta\right)}{\underset{j}{\prod}\ch^{2}\left(u_{j}-\eta\right)\underset{i<j}{\prod}\sh\left(v_{i}-v_{j}\right)}\det\hat{K}\Bigg\},\label{eq:S1S2_dynamic_matrix_element}
\end{multline}
where the matrix elements are
\begin{equation}
K_{ab}=T_{ab}+\frac{(-1)^{b}\sh^{3}\left(2\eta\right)\sh u_{b}\underset{l;l\neq b}{\prod}\sh\left(u_{l}-u_{b}+2\eta\right)}{\underset{i<j\neq b}{\prod}\sh\left(u_{i}-u_{j}\right)\underset{l;l\neq b}{\prod}\sh\left(u_{l}-u_{b}\right)}\allowdisplaybreaks
\frac{\sh\eta\underset{j,i;i\neq b}{\prod}\sh\left(u_{i}-v_{j}\right)\big[\frac{\sh u_{b}}{\ch\eta}+\underset{l}{\sum}\frac{\sh\left(2\eta\right)\ch\left(u_{b}+\eta\right)}{\ch\left(v_{l}-\eta\right)\ch\left(v_{l}+\eta\right)}\big]}{\ch\left(u_{b}+\eta\right)\ch\left(u_{b}+\eta\right)\ch\left(v_{a}-\eta\right)\ch\left(v_{a}+\eta\right)},\allowdisplaybreaks
\end{equation}
\begin{multline}
K_{ab}^{(y)}=T_{ab}+\frac{\left(-1\right)^{b}\sh^{3}\left(2\eta\right)\textrm{sgn}\left(y-b\right)\ch\left(u_{y}-\eta\right)}{\ch\left(v_{a}-\eta\right)\ch\left(v_{a}+\eta\right)\underset{i}{\prod}\sh\left(u_{b}-v_{i}\right)}\allowdisplaybreaks
\frac{\ch\left(u_{b}+\eta\right)\underset{l;l\neq y,b}{\prod}\sh\left(u_{l}-u_{b}+2\eta\right)}{\underset{i<j\neq y,b}{\prod}\sh\left(u_{i}-u_{j}\right)\underset{l;l\neq y,b}{\prod}\sh\left(u_{l}-u_{b}\right)}\\
\times\Big[\frac{\ch\left(u_{b}-\eta\right)}{\ch\left(u_{b}+\eta\right)}\allowdisplaybreaks
-\frac{\sh\left(u_{y}-u_{b}+2\eta\right)}{\sh\left(u_{y}-u_{b}-2\eta\right)}+\frac{\sh2\eta\ch\left(u_{b}-2\eta\right)\sh u_{y}}{\ch\left(u_{y}-\eta\right)\ch\left(u_{b}+\eta\right)}\Big],
\end{multline}
when $b\neq y$ and 
\begin{equation}
K_{ay}^{\left(y\right)}=\frac{\sh\left(2\eta\right)\sh\left(2v_{a}\right)}{\ch^{2}\left(v_{a}-\eta\right)ch^{2}\left(v_{a}+\eta\right)}, \label{eq:Kyay}
\end{equation}
$G_{ab}^{\left(y\right)}=T_{ab}$ when $b\neq y$ and $G_{ay}^{\left(y\right)}=K_{ay}^{\left(y\right)}$,
\begin{equation}
T_{ab}=\frac{\ch^{L}\left(v_{b}-\eta\right)}{\ch^{L}\left(v_{b}+\eta\right)}\frac{\sh\left(2\eta\right)}{\sh^{2}\left(v_{b}-u_{a}\right)}\prod_{j;j\neq a}\frac{\sh\left(v_{b}-u_{j}+2\eta\right)}{\sh\left(v_{b}-u_{j}\right)}\allowdisplaybreaks
-\frac{\sh\left(2\eta\right)}{\sh^{2}\left(u_{a}-v_{b}\right)}\prod_{j;j\neq a}\frac{\sh\left(u_{j}-v_{b}+2\eta\right)}{\sh\left(u_{j}-v_{b}\right)}.
\end{equation}
\end{widetext}
The normalisation factors of Bethe states\cite{Gaudin81_main,Korepin82_main} $\left\langle 0|0\right\rangle $
and $\left\langle \mathbf{q}|\mathbf{q}\right\rangle $ 
are quoted in Appendix B in terms of a determinant
of an $N\times N$ matrix. Here $\eta=(\textrm{acosh}1)/2$, 
\begin{equation}
u_{j}=\ln\left(\frac{\sqrt{1-e^{iq^0_{j}-2\eta}}}{\sqrt{1-e^{-iq^0_{j}-2\eta}}}\right)-i\frac{q^0_{j}}{2}
\end{equation}
are the quasimomenta of the ground  state $\mathbf{q}^0$ in Orbach
parametrisation, and $v_j$ is obtained from $u_j$ by $q^0_j\rightarrow q_j$ where $\mathbf{q}$ are the excited states. \cite{Gaudin_book}

\subsection{Hierarchy of modes}
\begin{figure}
	\begin{center}\includegraphics[width=1\columnwidth]{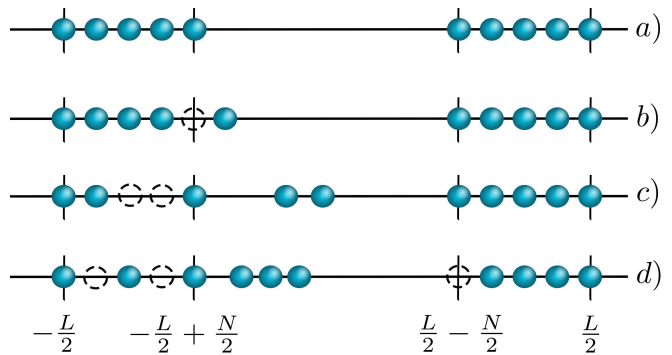}\end{center}\caption{\label{fig:quasimomenta_configs}Configurations of non-equal integer numbers $I_j$ that correspond to the solutions the Bethe equations for the model $H_m$: a) the ground state, b) one $\psi\psi^*$-pair excitation, c) two $\psi\psi^*$-pair excitations, b) three $\psi\psi^*$-pair excitations. These eigenstates include complex solutions at $\Delta=1$, which are obtained via numerical deformation of Bethe equations from the $\Delta=0$ to the $\Delta=1$ point.}
\end{figure}

The excitations in the sum over $\mathbf{q}$ in Eq.~(\ref{eq:dv2}) have the same number of quasimomenta as the ground state at a given magnetic field. They are constructed by removing a quasimomentum from the ground state distribution and promoting it to an empty position, see sketch in Fig. \ref{fig:quasimomenta_configs}. We will label these excitations as $\psi\psi^*$-pairs.

The whole dynamical correlation function in Eq. (\ref{eq:dv2}) exhibits
a hierarchy of modes governed by their spectral strength.
\cite{OT15,OT16, Moreno16} The excitations split into groups according to
$n=1$, $2$, $3$, $\dots$ $\psi\psi^*$-pairs that have progressively smaller amplitudes of their matrix
elements, $\left|\left\langle \mathbf{q}|\mathbf{S}_{1}\cdot\mathbf{S}_{2}|0\right\rangle \right|^{2}\sim1/L^{2n}$.
We keep the first three levels of the hierarchy, 
\begin{equation}
\delta v_{2}=\frac{J_{1}^{2}b^{2}\hbar}{mv_{0}J}(A_{1}+A_{2}+A_{3}).
\end{equation}
The first level consists of only one pair with the minimally possible
momentum $Q=p$, 
\begin{equation}
A_{1}\left(B\right)=\frac{v_m J L^{2}\left|\left\langle p|\mathbf{S}_{1}\cdot\mathbf{S}_{2}|0\right\rangle _{m}\right|^{2}}{2\pi\omega_{D}^{2}}, \label{eq:A1}
\end{equation}
where $v_{m}=\left(\varepsilon_{p}-\varepsilon_{0}\right)/p$ is the
renormalised velocity of Luttinger liquid and we have used smallness
of the exchange energy compared with Debye energy, $J/\omega_{D}\sim10^{-3}$
for general material parameters.\cite{Kittel_book} The only matrix
element in Eq. (\ref{eq:A1}) can be obtained using the bosonic modes
of Luttinger liquid, \cite{Giamarchi_book}
where the dispersion is almost linear. We, however, use a more general
Bethe ansatz approach here. Solutions of Bethe equations give $v_{m}$
directly that, together with the matrix element in Eq. (\ref{eq:S1S2_dynamic_matrix_element}),
gives the magnetic field dependence of $A_{1}\left(B\right)$ shown in the inset (right axis)
in Fig. \ref{fig:static_corr}. 

There are polynomially many states in the second and in the third
levels of the hierarchy,
\begin{equation}
A_{2\left(3\right)}\left(B\right)=\sum_{\left\{ \mathbf{q}\right\} ;Q=p}\frac{L\left|\left\langle \mathbf{q}|\mathbf{S}_{1}\cdot\mathbf{S}_{2}|0\right\rangle _{m}\right|^{2}}{\varepsilon_{\mathbf{q}}-\varepsilon_{0}},\label{eq:A23}
\end{equation}
where the summand in Eq. (\ref{eq:dv2}) was expanded in a Taylor
series in $\omega_{D}p/(\varepsilon_{0}-\varepsilon_{\boldsymbol{q}})\ll1$
since the sum over $\mathbf{q}$ accumulates dominantly at high energy\@.
Contribution of the low-energy excitations (for which $(\varepsilon_{0}-\varepsilon_{\boldsymbol{q}})/\omega_{D}p\ll1$)
has an additional small factor $J^{2}/\omega_{D}^{2}$, like in Eq.~(\ref{eq:A1}). At intermediate energies, $\left(\varepsilon_{0}-\varepsilon_{\boldsymbol{q}}\right)\simeq\omega_{D}p$,
the perturbation theory for $E\left(k\right)$ becomes inapplicable
since these magnetic excitations are in resonance with the acoustic
phonon. However, the width of the anti-crossing\cite{anticrossing} $\lesssim J_{1}\sqrt{\hbar b\omega_{D}^{2}/\left(mJ^{3}L^{5}\right)}$
is much smaller than the many-magnon level spacing
$J/L$ that is still in the Luttinger liquid regime. The non-perturbative
contribution of these levels is of the order of the anti-crossing width
and vanishes in large systems.

\begin{figure}
	\begin{center}\includegraphics[width=1\columnwidth]{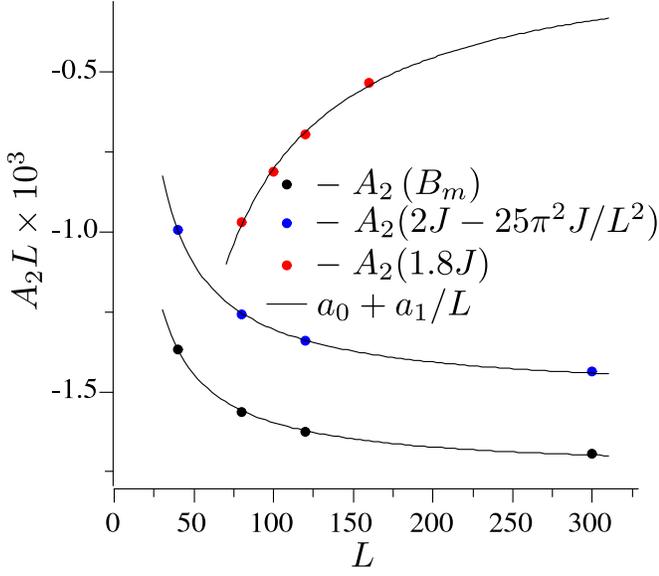}\end{center}\caption{\label{fig:A2_scaling}Scaling of $A_{2}L$ defined in Eq. (\ref{eq:A23}) with the system length at three values of the
		magnetic field $B=B_{m},\,2J-25\pi^{2}J/L^{2},\,1.8J$. The fitting
		of finite size corrections, $A_{2}L=a_{0}+a_{1}/L$, gives $\left(a_{0},a_{1}\right)\times 10^2=\left(-0.17,1.51\right),\,\left(-0.15,2.04\right),\,\left(-0.01,-6.95\right)$
		for the three magnetic fields respectively.}
\end{figure}

We obtain the magnetic field dependence of $A_{2}$ numerically as
a sum over the two $\psi\psi^*$-pairs in Eq.~(\ref{eq:A23}),
see inset in Fig. \ref{fig:static_corr}. At high fields $A_{2}$
is small since there are only a few excitations, the strength  of which is
small as $1/L^{4}$ at the second level of the hierarchy, and at small
fields $A_{2}$ is again small since the majority of the excitations belongs
to the class of pi-pairs close to the half-filling of the magnetic
band, which makes their amplitudes even weaker than $1/L^{4}$ due
to Eq. (\ref{eq:pi_pairs}). At the intermediate fields the $1/L^{4}$
smallness is partially compensated by a large number of the excitations,
whose majority does not have pi-pairs yet. The position of the maximum
of $\left|A_{2}\left(B\right)\right|$ is identified from numerics
at $B_{m}=2J-9\pi^{2}J/(2L^{2})$. The value of the function
at this point is $A_{2}\left(B_{m}\right)=-0.0016/L$ for large systems,
see scaling of $A_{2}\left(B_{m}\right)$ in Fig. \ref{fig:A2_scaling},
which is small in a different parameter compared with $A_{1}$. 

\begin{figure}
	\begin{center}\includegraphics[width=1\columnwidth]{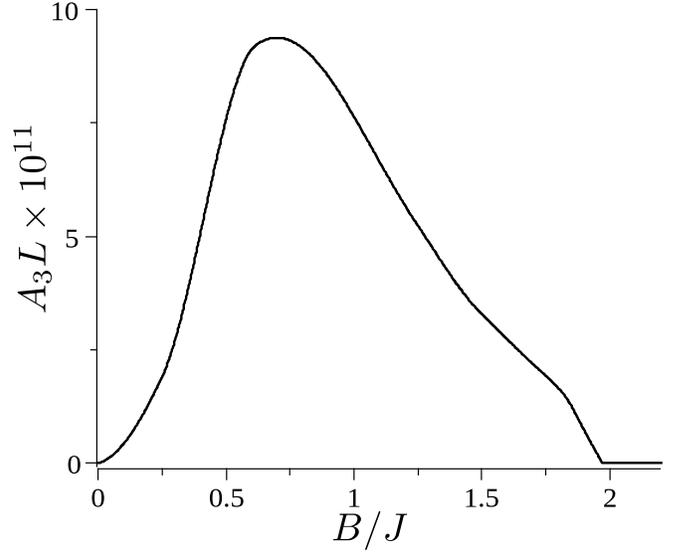}\end{center}\caption{\label{fig:A3}Contribution of the third level of the hierarchy of modes to $\delta v_2$ defined in Eq. (\ref{eq:A23}); $L=40$. It is small compared with $A_2$ in inset in Fig. \ref{fig:static_corr} for the whole range of magnetic fields.}
\end{figure}

For typical values of material parameters, $A_{1}$ and $A_{2}$ are of
the same order, \textit{e.g.} $1/L\sim10^{-6}$ and $\left(\omega_{D}/J\right)^{2}\sim10^{-6}$
for ultrasonic measurements in a magnetic insulator \cite{Luethi_book}.
The three $\psi\psi^*$-pairs contribution $A_{\text{3}}$ is smaller than
$A_{2}$ due to an additional $1/L^{2}$ in accord with the hierarchy
of modes\cite{OT15,OT16, Moreno16} for the whole range of magnetic fields,
see Fig. \ref{fig:A3}. 
\begin{figure}
\begin{center}\includegraphics[width=0.9\columnwidth]{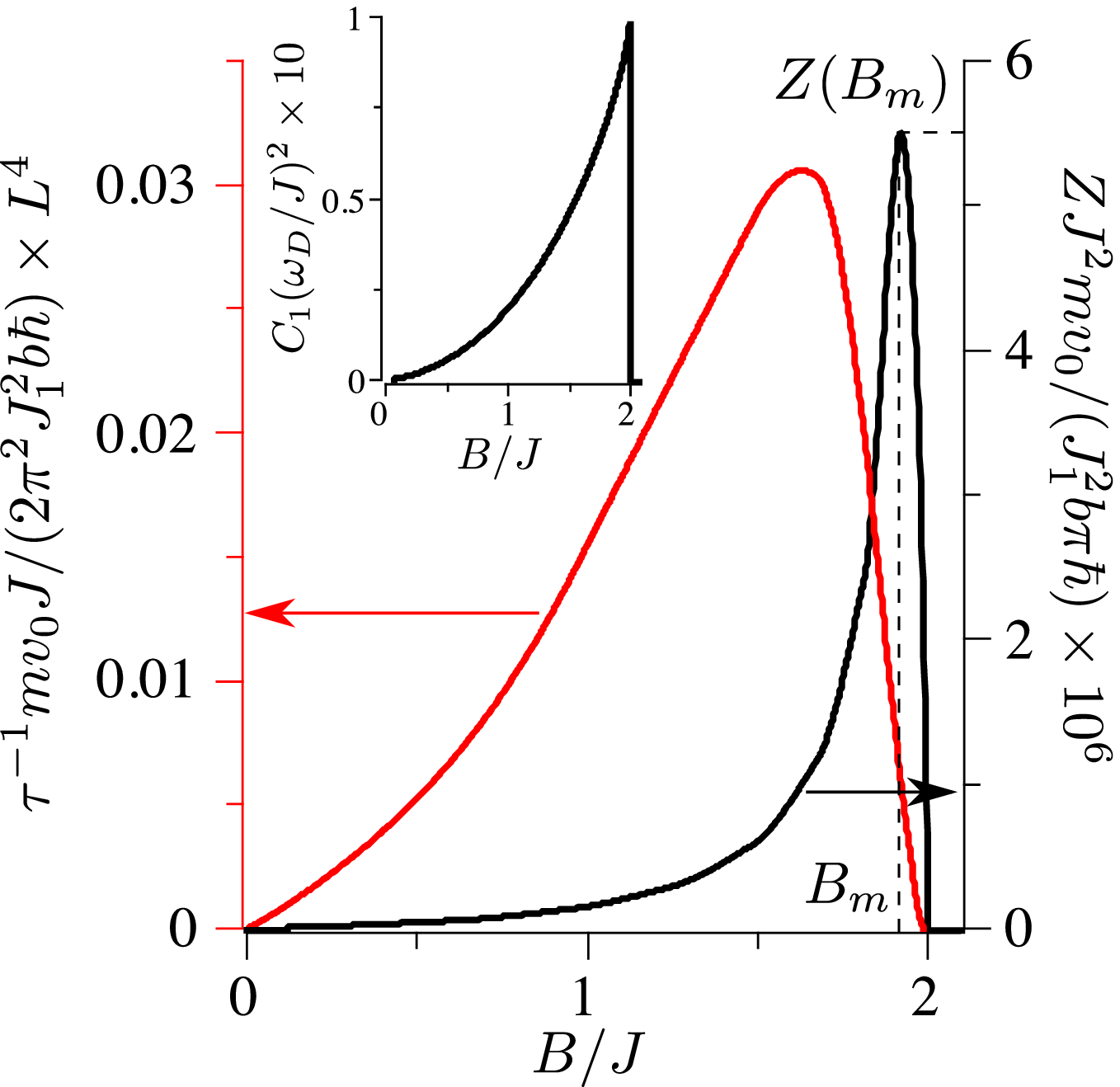}\end{center}\caption{\label{fig:tau}Two mechanisms of the sound attenuation: the red line
is the relaxation rate $\tau^{-1}$ calculated using the Fermi golden
rule in Eq. (\ref{eq:tau_Fermi}), while the black line is the degree
of hybridisation $Z$ of a sound phonon with the magnetic excitations 
dominated by high energies in Eq. (\ref{eq:I_hybridisation}). Inset is the low energy
contribution to Eq. (\ref{eq:I_hybridisation}).}
\end{figure}

\section{Attenuation of phonons}
Next let us analyse decay of the phonons into the spin excitations. The
excitation spectrum of Heisenberg model in Eq. (\ref{eq:H_m_ph})
is continuous which always has some states in resonance with the single
phonon energy $\omega_{D}p$ providing a channel for the direct relaxation,
unlike the previous phenomenological approaches. \cite{Kreisel11,Streib15}
The rate of such a process is given by Fermi golden rule,
\begin{multline}
\tau^{-1}=\frac{2\pi^{2}J_{1}^{2}b}{mv_{0}}\sum_{\left\{ \mathbf{q}\right\} ;Q=p}\left|\left\langle \mathbf{q}|\mathbf{S}_{1}\cdot\mathbf{S}_{2}|0\right\rangle _{m}\right|^{2}\delta\left(\Delta E\right),\label{eq:tau_Fermi}
\end{multline}
where $\Delta E=\varepsilon_{\mathbf{q}}-\varepsilon_{0}-\omega_{p}$
and the contribution of the $J_{2}$ term in Eq. (\ref{eq:V}) is zero
due to $\delta\left(\Delta E\right)$. The principal value of the
sum in Eq. (\ref{eq:tau_Fermi}) is accumulated by the second level
of the hierarchy, which we evaluate numerically \textendash{} see
the magnetic field dependence of $\tau^{-1}$ in Fig. \ref{fig:tau}.
Its maximum value has the same small prefactor $1/L^{4}$ as the matrix
element in Eq. (\ref{eq:S1S2_dynamic_matrix_element}) making the
direct relaxation extremely slow in large systems. 

\begin{figure}
	\begin{center}\includegraphics[width=1\columnwidth]{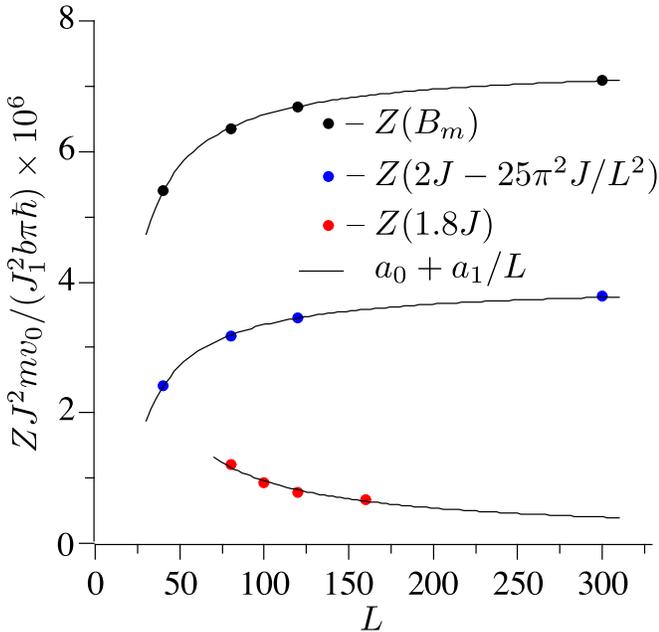}\end{center}\caption{\label{fig:Z_scaling}Scaling of $Z$ defined in Eq. (\ref{eq:I_hybridisation}) with the system length at three values of the magnetic
		field $B=B_{m},\,2J-25\pi^{2}J/L^{2},\,1.8J$. The fitting of finite
		size corrections, $ZJ^{2}mv_{0}/\left(J_{1}^{2}b\pi\hbar^{2}\right)=a_{0}+a_{1}/L$,
		gives $\left(a_{0},a_{1}\right)\times10^{5}=\left(0.73,-7.8\right)$,
		$\left(0.40,-6.3\right)$, $\left(0.012,8.4\right)$ for the three
		magnetic fields respectively.}
\end{figure}

\begin{figure}
	\begin{center}\includegraphics[width=1\columnwidth]{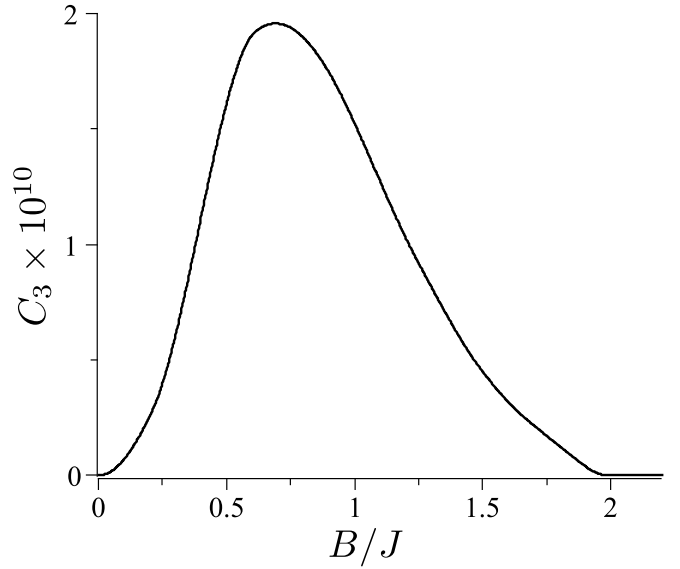}\end{center}\caption{\label{fig:C3}Contribution of the third level of the hierarchy of modes to $Z$ defined in Eq. (\ref{eq:C3}); $L=40$. It is small compared with $C_2$ in  Fig. \ref{fig:tau} for the whole range of magnetic fields.}
\end{figure}

However, the amplitude of the free phonons can also be reduced via
hybridisation with the magnetic excitations, similarly to the $\delta v_{2}$
renormalisation of their velocity. The first order in perturbation
theory for the wave function, 
\begin{equation}
\left|\Psi_{k}\right\rangle =\left|k\right\rangle +\sum_{\left\{ \mathbf{k},\mathbf{q}\right\} }\frac{\left\langle \mathbf{k},\mathbf{q}|V|k\right\rangle }{\varepsilon_{0}+\omega_{k}-\varepsilon_{\mathbf{q}}-\omega_{\mathbf{k}}}\left|\mathbf{k}\right\rangle _{ph}\left|\mathbf{q}\right\rangle _{m},
\end{equation}
gives suppression at low momenta, $Z=1-\left|\left\langle p|\Psi_{p}\right\rangle \right|^{2}$,
as 

\begin{equation}
Z=\frac{J_{1}^{2}b\pi\hbar^{2}}{mv_{0}}\sum_{\left\{ \mathbf{q}\right\} ;Q=p}\frac{|\left\langle \mathbf{q}|\mathbf{S}_{1}\cdot\mathbf{S}_{2}|0\right\rangle _{m}|^{2}}{\left(\varepsilon_{0}-\varepsilon_{\mathbf{q}}-\omega_{D}p\right)^{2}},\label{eq:I_hybridisation}
\end{equation}
which we analyse using the hierarchy of modes: 
\begin{equation}
Z=\frac{J_{1}^{2}b\pi\hbar^{2}}{J^{2}mv_{0}}(C_{1}+C_{2}+C_{3}),
\end{equation}
where 
\begin{equation}
C_{1}=\frac{(J/\omega_{D})^{2}L^{2}|\left\langle p|\mathbf{S}_{1}\cdot\mathbf{S}_{2}|0\right\rangle _{m}|^{2}}{(2\pi)^{2}}
\end{equation}
and 
\begin{equation}
C_{2\left(3\right)}=\sum_{\left\{ \mathbf{q}\right\} ;Q=p}\frac{|\left\langle \mathbf{q}|\mathbf{S}_{1}\cdot\mathbf{S}_{2}|0\right\rangle _{m}|^{2}}{(\varepsilon_{0}-\varepsilon_{\mathbf{q}})^{2}},\label{eq:C3}
\end{equation}
like in the analysis of Eq. (\ref{eq:dv2}) before. The first level
contribution $C_{1}$ is small
in $J^{2}/\omega_{D}^{2}$ like $A_{1}$, see inset in Fig. \ref{fig:tau}. But $C_{2}$, shown Fig.
\ref{fig:tau}, remains finite in the thermodynamic limit  (see scaling
in Fig. \ref{fig:Z_scaling}) unlike $A_{2}$ above, and $C_{3}$
is small in $1/L^{2}$ compared with $C_{2}$, see Fig. \ref{fig:C3}. 

This hybridisation
mechanism can be distinguished from other non-magnetic channels of
relaxation via its magnetic field dependence and from the exponential
decay into the resonant magnetic states described by Eq. (\ref{eq:tau_Fermi})
since it is constant in the temporal and spatial domains.

\section{Ultrasound experiment on $\textrm{Cs}_{2}\textrm{CuCl}_{4}$}
Finally, we discuss our experimental results. High-quality single crystals of several mm size of the frustrated spin-1/2 antiferromagnet  $\textrm{Cs}_{2}\textrm{CuCl}_{4}$ were grown from an aqueous solution by an evaporation technique.\cite{Kruger10} A pair of piezoelectric polymer-foil transducers was glued to opposite parallel surfaces perpendicular to the [010] direction for the generation and the detection of the ultrasound waves. These longitudinal waves  propagate along the [010] direction that corresponds to the elastic mode $c_{22}$. Changes of the sound velocity $\delta v$ and the renormalised amplitude of the sound wave $1-Z$ were measured as functions of magnetic field at constant temperatures, using the experimental set up described in detail in Ref. \onlinecite{Luthi94}.

\begin{figure}
	\begin{center}\includegraphics[width=1\columnwidth]{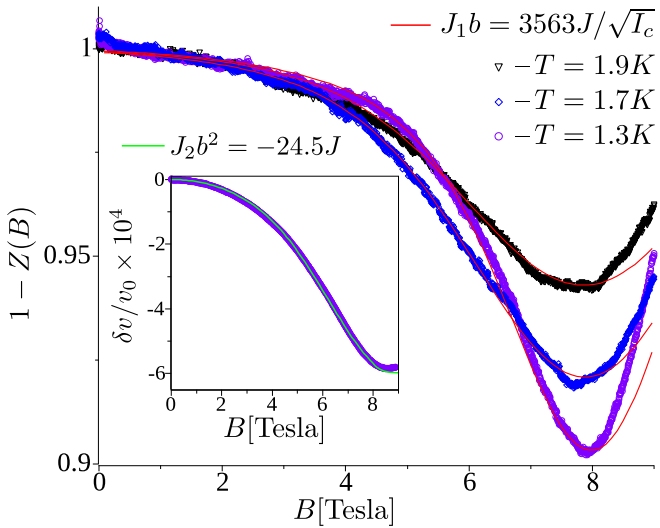}\end{center}\caption{\label{fig:exp_fits} Experimental results (open triangles, diamonds, and circles) of the renormalised amplitude $1-Z$ of the longitudinal ultrasound wave propagating along the [010] axis of $\textrm{Cs}_{2}\textrm{CuCl}_{4}$ at $T = (1.900\pm0.005)$ K, $T = (1.700\pm0.005)$ K, and $T = (1.300\pm0.005)$ K. The red lines represent the results of Eq. (\ref{eq:I_hybridisation}) with $J_{1}b=3563J/\sqrt{I_c}$. The inset shows data (open circles) of the corresponding normalised sound velocity at $T = (1.300\pm0.005)$ K. The green line shows the result of Eq. (\ref{eq:dv1}) using $J_{2}b^{2}=-24.5J$. Additional data are presented in Fig. \ref{fig:exp_additional}.}
\end{figure}
In Fig. \ref{fig:exp_fits} we compare  the experimental data for the sound velocity with the theoretical results expressed in Eqs.~(\ref{eq:dv1},\ref{eq:dv2}). By fitting the static correlation function given by Eq.~(\ref{eq:dv1}), we extract $J_2b^2 = -24.5J$, with the magnetic coupling constant $J=0.375$ meV taken from \cite{Coldea03} \textendash{} see inset in Fig. \ref{fig:exp_fits}. We find no signatures of the dynamical correlation functions represented by Eq. (\ref{eq:dv2}) \textendash{} which are parametrically small \textendash{} down to the noise level of our experiment. This defines an upper bound to the other microscopic constant $J_{1}b\leq1.25\times 10^{4}J$. 

Analysing the attenuation of the amplitude of the sound wave $Z$, we find that its functional dependence on the magnetic field is in good agreement with the dynamic hybridisation mechanism represented by Eq. (\ref{eq:I_hybridisation}) \textendash{} see Fig. \ref{fig:exp_fits} and additional data in Fig. \ref{fig:exp_additional}. By fitting its amplitude, we extract the other microscopic parameter as 
$J_{1}b=3563J/\sqrt{I_{c}}$, where $I_{c}$ is the
degree of non-magnetic losses. A quantitative determination of these losses is not possible since they consist of various extrinsic (\emph{e.g.} coupling and diffraction losses, non-parallel alignment of the sample, \emph{etc}) and intrinsic attenuation mechanisms like direct scattering at defects or dislocation damping. \cite{Truell_book} However, even for $I_{c}=1$ this value of $J_{1}b$ is consistent
with the bound from the measurement of the sound velocity. 

The values of the microscopic constants are significantly different from the values measured along the $a$-axis in Ref. \onlinecite{Wolf16} manifesting an anisotropy of $\textrm{Cs}_{2}\textrm{CuCl}_{4}$. Our very good fit of the magnetic field dependencies by the purely one-dimensional theory in Figs. \ref{fig:exp_fits} and \ref{fig:exp_additional} gives a further argument that the inter-chain interactions in $\textrm{Cs}_{2}\textrm{CuCl}_{4}$ in the finite temperature regime are negligible despite only a moderate degree of the exchange  anisotropy of $\sim 3$ in the $a-b$ plane. \cite{Balents10}

\section{Conclusions}
In conclusion, constructing a microscopic theory of magneto-elasticity
in 1D we have shown that the necessary correlation functions involve
the many-body excitations at all energy scales and have identified
a new mechanism of sound attenuation. Our theoretical predictions
agree  with our ultrasound experiments in the 1D regime of
$\textrm{Cs}_{2}\textrm{CuCl}_{4}$.

\section{Acknowledgements}
We acknowledge financial support by the DFG through the SFB/TRR
49. 

\appendix
\begin{figure*}[t]
	\begin{center}\includegraphics[width=1\textwidth]{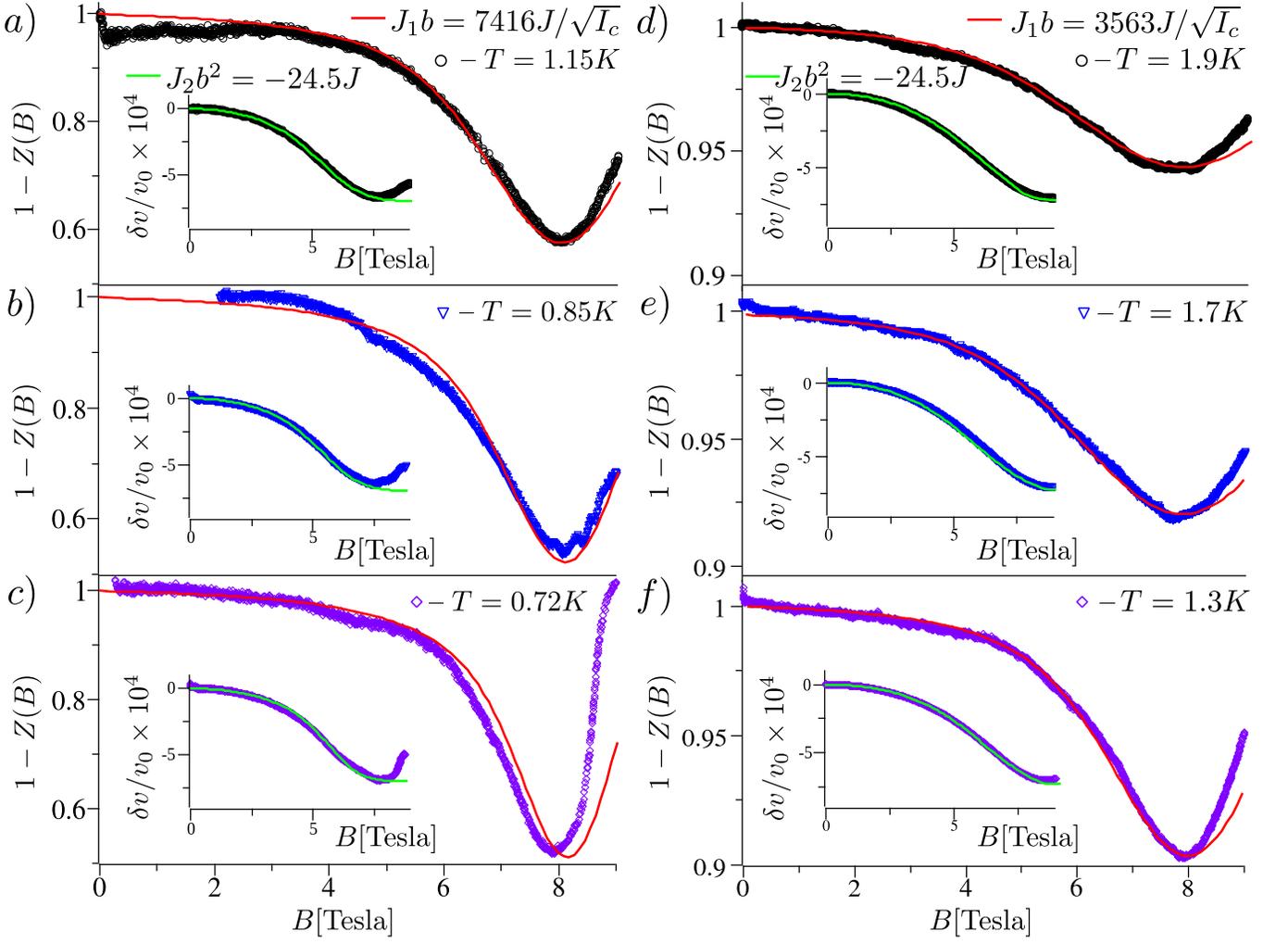}\end{center}\caption{\label{fig:exp_additional}Experimental results (open circles, triangles, and diamonds) of the renormalised amplitude $1-Z$ of the longitudinal ultrasound wave propagating along [010] axis of $\textrm{Cs}_{2}\textrm{CuCl}_{4}$ at the temperatures of a) $T=(1.150\pm0.005)$ K, b) $T=(0.850\pm0.005)$ K, c) $T=(0.720\pm0.005)$ K, d) $T=(1.900\pm0.005)$ K, e) $T=1.700\pm0.005)$ K, and f) $T=(1.300\pm0.005)$ K.  The red lines represent the results of Eq.~(\ref{eq:I_hybridisation}) with $J_1 b=7416J/\sqrt{I_c}$ for a,b,c) and with $J_1b=3563/\sqrt{I_c}$ for d,e,f). The data in d,e,f) were taken in a different cryostat system using upgraded electronics and a better quality sample compared with the data in a,b,c) leading to a decrease of the non-magnetic losses by a factor of $I_c(a,b,c)/I_c(d,e,f)\approx 4.3$. The insets show data of the corresponding normalised sound velocities at the same temperatures. The green lines show the result of Eq. (\ref{eq:dv1}) using $J_2b^2=-24.5J$. The results of our ultrasound experiments are still relatively close to the 1D regime at the temperature of $T=(0.720\pm0.005)$ K, at which the system is already in a transitional state between the 1D and a 2D regime.}
\end{figure*}

\section{Derivation of the quantisation equation for pi-pairs}

The XXZ model is a generalisation of Heisenberg model that introduces
the interaction strength between magnons $J\Delta$ as a model parameter,
which breaks the rotational symmetry of the spin-spin interaction
term $\mathbf{S}_{j}\cdot\mathbf{S}_{j+1}$. In one-dimension the
model reads

\begin{equation}
H_{m}=\sum_{j=1}^{L}\Big(J\frac{S_{j}^{-}S_{j+1}^{+}+S_{j}^{+}S_{j+1}^{-}}{2}+J\Delta S_{j}^{z}S_{j+1}^{z}+BS_{j}^{z}\Big),\label{eq:Hm_app}
\end{equation}
where $S_{j}^{\pm}=S_{j}^{x}\pm iS_{j}^{y}$. For $\Delta=1$ this
model becomes the model in Eq. (2). The $N$-magnon eigenstates of this Hamiltonian can be
found by solving a set of Bethe equations,

\begin{equation}
q_{j}L-\sum_{l\neq j}\varphi_{jl}=2\pi I_{j},\label{eq:BA_app}
\end{equation}
for $N$ quasimomenta $q_{j}$, where the two-magnon scattering phases
are given by 
\begin{equation}
e^{i\varphi_{ij}}=-\frac{e^{i\left(q_{i}+q_{j}\right)}+1-2\Delta e^{iq_{i}}}{e^{i\left(q_{i}+q_{j}\right)}+1-2\Delta e^{iq_{j}}}\label{eq:phi_general_app}
\end{equation}
and $I_{j}$ is a set of $N$ non-equal integer numbers.

In the free magnon limit $\Delta=0$ the two-body phase shifts $\varphi_{ij}$
become  independent of quasimomenta and equal to the shifts of free fermions
or hard-core bosons, $\varphi_{ij}=\pi$ that is immediately obtained
by taking the $\Delta\rightarrow0$ limit of Eq. (\ref{eq:phi_general_app})
giving $\exp\left(i\varphi_{ij}\right)=-1$. This results restores
the plain wave quantisation condition for each magnon independently,
$q_{j}=2\pi\left(I_{j}+1/2\right)/L$, \cite{parity_note} from the system of non-linear Bethe equations in Eq.~(\ref{eq:BA_app}).
Alternatively, the free magnon result can be obtained by setting $\Delta=0$
in the Hamiltonian in Eq.~(\ref{eq:Hm_app}) and then by diagonalising
the resulting XY model directly. \cite{Lieb61}

However, the non-interacting limit becomes ambiguous when at least
one pair of quasimomenta in an $N$ magnon state satisfies the condition
$q_{i}+q_{j}=2\pi\left(I_{i}+I_{j}+1\right)/L=\pm\pi$ at $\Delta=0$.
In evaluating the $\Delta=0$ limit of Eq. (\ref{eq:phi_general_app}),
the leading order of $e^{i\left(q_{i}+q_{j}\right)}+1$ is zero in
the Taylor series around the $\Delta=0$ point, both in the numerator
and in the denominator. Thus, higher order coefficients have to be
calculated, $e^{i\left(q_{i}+q_{j}\right)}+1=c_{1}\Delta+c_{2}\Delta^{2}+\dots$,
that requires, in general, solving the whole set of $N$ Bethe equation
in Eq. (\ref{eq:BA_app}) in a non-linear fashion, i.e. $c_{1},c_{2},...$
depend on all $q_{l}$ \textendash{} not just $q_{i}$ and $q_{j}$
\textendash{} since $\Delta$ is still finite, which requires solving
all $N$ Bethe equations for all $q_{l}$ simultaneously in taking
the limit. This issue was noted in Refs. \onlinecite{McCoy01,Noh00} but was never
resolved. Here we start from a finite but small $\Delta$, for which
all scattering phase are well-behaved, and then take the limit $\Delta\rightarrow0$
systematically. 

Let us consider a $N$-magnon solution of Bethe equations that containts
$2r$ quasimomenta that satisfy the $q_{2j}+q_{2j+1}=\pm\pi$ condition
(we will call these pairs of quasimomenta pi-pairs below) and $N-2r$
quasimomenta that do not have a pi-pair. For a finite but small
$\Delta\ll1$ the quasimomenta of a pi-pair can be parameterised
as 
\begin{align}
q_{2j} & =\pm\pi-\alpha_{j}+\frac{\delta_{j}}{2},\qquad j\leq r,\label{eq:pi_pair_1}\\
q_{2j+1} & =\alpha_{j}+\frac{\delta_{j}}{2},\label{eq:pi_pair_2}
\end{align}
where $\alpha_{j}$ is an unknown parameter of the $j^{th}$ pi-pair
that has a non-zero value, since Bethe equations for this pair can
not be solved due to the unknown (at the moment) phase shift $\varphi_{2j,2j+1}$
at $\Delta=0$, and $\delta_{j}$ is a parameter that vanishes at
$\Delta=0$. On the other hand, the remaining quasimomenta $j>2r$
can be found immediately for $\Delta=0$ since all of their scattering
phases in Eq. (\ref{eq:phi_general_app}) for these quasimomenta are
well-behaved, $\varphi_{ij}=\pi$. Thus at a finite $\Delta\ll1$
we can write 
\begin{equation}
q_{j}=\frac{2\pi\left(I_{j}+\frac{1}{2}\right)}{L}+\epsilon_{j},\qquad j>2r,\label{eq:non_pi_q}
\end{equation}
where $\epsilon_{j}$ are small corrections due to a finite $\Delta$
that depend on all other quasimomenta and vanish for $\Delta=0$.
Conservation of the total momentum of $N$ magnons, $\sum_{j=1}^{N}q_{j}=2\pi\sum_{j=1}^{N}I_{j}/L$
including the pi-pairs, is independent of the interactions and imposes
an additional constraint on $\delta_{j}$ and $\epsilon_{j}$, 
\begin{equation}
\sum_{j=2r+1}^{N}\epsilon_{j}=-\sum_{j=1}^{r}\delta_{j}.\label{eq:Pm_contraint}
\end{equation}
It is obtained as a sum of all equation in Eq. (\ref{eq:BA_app})
after substitution of Eqs. (\ref{eq:pi_pair_1}-\ref{eq:non_pi_q}).

Since $\alpha_{j}$ cannot be obtained directly from its own Bethe
equation due to the undefined scattering phase within the corresponding
pi-pair, we are going to obtain an equation for $\alpha_{j}$ from
the other $j>2r$ Bethe equations that do not have this issue. We
start from expanding $\varphi_{ji}$ for $j>2r$ magnons \textendash{}
which are defined at the point $\Delta=0$ \textendash{} up to the
linear order in small $\Delta$. Taking into account that $\epsilon_{j}$
is proportional to $\Delta$ and $\delta_{j}$ is linear (or a higher
order) in $\Delta$ we obtain the following expansion of $\varphi_{ji}$
between the $j^{th}$ magnon and a pi-pair and the $j^{th}$ magnon
and another $j'^{th}$ magnons, which do not have a pi-pair,
respectively,

\begin{align}
\varphi_{j,2i} & =\pi-2\Delta\frac{\sin\frac{q_{j}-\alpha_{i}}{2}}{\cos\frac{q_{j}+\alpha_{i}}{2}},\label{eq:expansion_phi_2i}\\
\varphi_{j,2i+1} & =\pi-2\Delta\frac{\cos\frac{q_{j}+\alpha_{i}}{2}}{\sin\frac{q_{j}-\alpha_{i}}{2}},\label{eq:expansion_phi_2i+1}\\
\varphi_{jj'} & =\pi-2\Delta\frac{\sin\frac{q_{j}-q_{j'}}{2}}{\cos\frac{q_{j}+q_{j'}}{2}}.\label{eq:expansion_phi_jjp}
\end{align}
Then we substitute these expansions in Eq. (\ref{eq:BA_app}) and
sum all of them with $j>2r$ obtaining a relation between $\delta_{i}$
and $\alpha_{i}$ that parameterise the quasimomenta for $j\leq2r$,
\begin{equation}
-L\sum_{i=1}^{r}\delta_{i}+4\Delta\sum_{i=1}^{r}\sum_{j=2n+1}^{N}\frac{1-\sin q{}_{j}\sin\alpha_{i}}{\sin q_{j}-\sin\alpha_{i}}=0,\label{eq:sum_BA_jg2n}
\end{equation}
where the sum over $j$ is taken over the remaining quasimomenta that
do not have a pi-pair, given by Eq. (\ref{eq:non_pi_q}) with
$\epsilon_{j}=0$. Here we used Eq. (\ref{eq:Pm_contraint}) to express
$\epsilon_{j}$ through $\delta_{j}$ and cancelled all $\varphi_{jj'}$
terms for both $j,j'>2r$ due to the $\varphi_{jj'}=-\varphi_{j'j}$
symmetry of Eq. (\ref{eq:expansion_phi_jjp}) \textendash{} note that
the scattering phases are defined up to an arbitrary period $2\pi$
times an integer.

The equation in Eq. (\ref{eq:sum_BA_jg2n}) is a sum of $r$ terms
and each term depends only on two unknown variables $\delta_{i}$
and $\alpha_{i}$. Thus Eq. (\ref{eq:sum_BA_jg2n}) splits into $r$
independent equations and solving them separately we find 
\begin{equation}
\delta_{i}=\frac{4}{L}\sum_{j=2r+1}^{N}\frac{1-\sin q{}_{j}\sin\alpha_{i}}{\sin q_{j}-\sin\alpha_{i}}\Delta.\label{eq:delta_i}
\end{equation}
This result shows that the linear term in the Taylor expansion for
$\delta_{i}$ in $\Delta$ does not vanish. However it depends on
the still unknown parameter $\alpha_{i}$. In order to find it, we take
the $\Delta\rightarrow0$ limit of Eq. (\ref{eq:phi_general_app})
for the two quasimomenta within the $i^{th}$ pi-pair and obtain 
\begin{equation}
e^{i\varphi_{2i,2i+1}}=\frac{i\frac{2}{L}\sum_{j=2r+1}^{N}\frac{1-\sin q_{j}\sin\alpha_{i}}{\sin q_{j}-\sin\alpha_{i}}-e^{-i\alpha_{i}}}{i\frac{2}{L}\sum_{j=2r+1}^{N}\frac{1-\sin q{}_{j}\sin\alpha_{i}}{\sin q_{j}-\sin\alpha_{i}}+e^{i\alpha_{i}}},\label{eq:expansion_phi_pair}
\end{equation}
where 
\begin{equation}
e^{i\left(q_{2i}+q_{2i+1}\right)}+1=\frac{4}{L}\sum_{j=2r+1}^{N}\frac{1-\sin q{}_{j}\sin\alpha_{i}}{\sin q_{j}-\sin\alpha_{i}}\Delta
\end{equation}
was expanded up to the linear order in $\Delta$, substituted in Eq.
(\ref{eq:phi_general_app}), and $\Delta$ was cancelled from the whole
expression altogether. Substituting Eq. (\ref{eq:expansion_phi_pair})
into each of the $2i^{th}$ (or $2i+1^{st}$) Bethe equation in Eq.
(\ref{eq:BA_app}) in the exponential form we obtain an equation for
each $\alpha_{i}$ independently in the $\Delta=0$ limit,

\begin{equation}
-e^{iL\alpha_{i}}\frac{i\frac{2}{L}\sum_{j=2r+1}^{N}\frac{1-\sin q_{j}\sin\alpha_{i}}{\sin q_{j}-\sin\alpha_{i}}-e^{-i\alpha_{i}}}{i\frac{2}{L}\sum_{j=2r+1}^{N}\frac{1-\sin q{}_{j}\sin\alpha_{i}}{\sin q_{j}-\sin\alpha_{i}}+e^{i\alpha_{i}}}=1.\label{eq:BA_qi}
\end{equation}

This result does not coincide with the free wave quantisation condition
$\exp\left(i\alpha_{i}L\right)=-1$, being a non-linear equation for
$\alpha_{i}$. Its solutions can be found by introducing an extra
deformation parameter $\lambda$,
\begin{equation}
-e^{iL\alpha}\frac{i\frac{2\lambda}{L}\sum_{j=2r+1}^{N}\frac{1-\sin q_{j}\sin\alpha}{\sin q_{j}-\sin\alpha}-e^{-i\alpha}}{i\frac{2\lambda}{L}\sum_{j=2r+1}^{N}\frac{1-\sin q{}_{j}\sin\alpha}{\sin q_{j}-\sin\alpha}+e^{i\alpha}}=1,\label{eq:BA_q}
\end{equation}
where the subscript was omitted, $\alpha_{i}\rightarrow\alpha$, since
the equation is the same for all indices $i$. The solutions can be classified
in the limit $\lambda=0$, like the Bethe equations, where Eq.~ (\ref{eq:BA_q})
is solved by $\alpha=2\pi\left(I_{j}+1/2\right)/\left(L-2\right)$.
Then a smooth deformation of the equation from $\lambda=0$ to $\lambda=1$
gives all solutions of of the non-linear Eq. (\ref{eq:BA_qi}). 
The quantisation equation of pi-pairs in the $\Delta=0$ limit before Eq. (\ref{eq:pi_pairs}) is Eq. (\ref{eq:BA_q}) in the logarithmic form.

The two-magnon solutions of Bethe equations that we identified as
pi-pairs in Eq. (\ref{eq:BA_q}) at $\Delta=0$ can account for the
missing complex solutions, which instead remain real, of the XXX model
at $\Delta=1$ found in Ref. \onlinecite{Essler92}. For $N=2$ the parameter $\delta$
in Eq.~(\ref{eq:delta_i}) remains zero for any $\Delta$ making the
scattering phase $\varphi_{12}=-2\alpha$ and Eq. (\ref{eq:BA_app})
independent of $\Delta$ as well, in this case. Thus this two-magnon
solution remains real at $\Delta=1$ and has to be removed from the
class of complex conjugated quasimomenta. We also note that pi-pairs
are still solutions of Bethe equations at any finite $\Delta$ in
full accord with the arguments of Ref. \onlinecite{Baxter02}. It is only the
limit $\Delta\rightarrow0$ of these solutions that does not recover
the single particle quantitation rule $q_{j}=2\pi\left(I_{j}+1/2\right)/L$.

\section{Normalisation factors of Bethe states}
\setcounter{equation}{0}

The eigenstates of the XXZ model in Eq.~(\ref{eq:Hm_app}) are the
Bethe states 
\begin{equation}
\left|\Psi\right\rangle =\sum_{\mathcal{P},j_{1}<\dots<j_{N}}\allowbreak e^{i\sum_{l}q_{\mathcal{P}_{l}}j_{l}+i\sum_{l<l'}\varphi_{\mathcal{P}_{l},\mathcal{P}_{l'}}/2}S_{j_{1}}^{+}\dots S_{j_{N}}^{+}\left|\Downarrow\right\rangle,
\end{equation}
where $\mathcal{P}$ is a permutation of $N$ quasimomenta $q_{j}$
and $\left|\Downarrow\right\rangle$ is the ferromagnetic ground
state. In this so-called coordinate representation the many-body states $\left|\Psi\right\rangle $
are not factorisable making calculations of scalar products and expectation
values in this representation almost intractable. However, a calculation
of the form factors needed in Eq. (13) becomes manageable
using the algebraic form of Bethe ansatz, \cite{Korepin_book_main} in
which Bethe states are factorised in terms of operators with given
commutation relations. 

Following Ref. \onlinecite{Korepin_book_main} we write down the many-body
wave functions using operators that satisfy a Yang-Baxter algebra
as 

\begin{equation}
\left|\mathbf{u}\right\rangle =\prod_{j=1}^{N}C\left(u_{j}\right)\left|\Downarrow\right\rangle ,\label{eq:psi_ABA}
\end{equation}
where $u_{j}$ are $N$ auxiliary parameters and $C\left(u\right)$
is one of the four matrix elements of the transition matrix 
\begin{equation}
T\left(u\right)=\left(\begin{array}{cc}
A\left(u\right) & B\left(u\right)\\
C\left(u\right) & D\left(u\right)
\end{array}\right),\label{eq:BA_operators}
\end{equation}
which is defined in an auxiliary $2\times 2$ space. This $T$-matrix satisfies
the Yang-Baxter equation
\begin{equation}
R\left(u-v\right)\left(T\left(u\right)\otimes T\left(v\right)\right)=\left(T\left(v\right)\otimes T\left(u\right)\right)R\left(u-v\right).\label{eq:YB_equation}
\end{equation}
Here we use the following $R$-matrix that corresponds to the spin
Hamiltonian in Eq.(\ref{eq:Hm_app}), 
\begin{equation}
R\left(u\right)=\left(\begin{array}{cccc}
1\\
& b\left(u\right) & c\left(u\right)\\
& c\left(u\right) & b\left(u\right)\\
&  &  & 1
\end{array}\right),\label{eq:R_matrix}
\end{equation}
where $b\left(u\right)=\sinh\left(u\right)/\sinh\left(u+2\eta\right)$
and $c\left(u\right)=\sinh\left(2\eta\right)/\sinh\left(u+2\eta\right)$.

The entries of Eq. (\ref{eq:YB_equation}) give commutation relations
between the matrix elements of $T$. Here we write down four of them
that will be used later,

\begin{equation}
\left[B_{u},C_{v}\right]=\frac{c\left(u-v\right)}{b\left(u-v\right)}\left(A_{u}D_{v}-A_{v}D_{u}\right),\label{eq:ABA_BC}
\end{equation}
\begin{equation}
A_{u}C_{v}=\frac{1}{b\left(u-v\right)}C_{v}A_{u}-\frac{c\left(u-v\right)}{b\left(u-v\right)}C_{u}A_{v},\label{eq:ABA_AC}
\end{equation}
\begin{equation}
D_{u}C_{v}=\frac{1}{b\left(v-u\right)}C_{v}D_{u}-\frac{c\left(v-u\right)}{b\left(v-u\right)}C_{u}D_{v},\label{eq:ABA_DC}
\end{equation}
\begin{equation}
\left[A_{u},D_{v}\right]=\frac{c\left(u-v\right)}{b\left(u-v\right)}\left(C_{v}B_{u}-C_{u}B_{v}\right).\label{eq:ABA_AD}
\end{equation}
We have introduced the subscript $u$ and $v$ as a shorthand of the
argument, e.g. $A_{u}\equiv A\left(u\right)$, above.

The transfer matrix $\tau\left(u\right)=\textrm{Tr}T\left(u\right)=A\left(u\right)+D\left(u\right)$
contains all of the conserved quantities of the model in Eq. (\ref{eq:Hm_app})
including the Hamiltonian. Thus if $\left|\mathbf{u}\right\rangle $
is a eigenstate of $\tau\left(u\right)$ then it is an eigenstate
of the Hamiltonian. The eigenvalue equation, $\tau\left(u\right)\left|\mathbf{u}\right\rangle =\mathcal{T}_{u}\left|\mathbf{u}\right\rangle $
where $\mathcal{T}_{u}$ is a scalar quantity \textendash{} the corresponding
eigenvalue, can be solved using the commutation relations in Eqs.~
(\ref{eq:ABA_BC}-\ref{eq:ABA_AD}). The results of acting with the
$A_{u}$ and $D_{u}$ operators on the state $\left|\mathbf{u}\right\rangle $
in Eq.~(\ref{eq:psi_ABA}) are obtained by commuting them from left
to right through the product of $C\left(u_{j}\right)$ operators,
\begin{widetext}
	
	\begin{equation}
	A_{u}\prod_{j=1}^{N}C\left(u_{j}\right)\left|0\right\rangle =a_{u}\prod_{j=1}^{N}\frac{1}{b_{uj}}C\left(u_{j}\right)\left|0\right\rangle -\sum_{j=1}^{N}a_{j}\frac{c_{uj}}{b_{uj}}C\left(u\right)\prod_{l=1\neq j}^{N}\frac{1}{b_{jl}}C\left(u_{l}\right)\left|\Downarrow\right\rangle ,\label{eq:A_string_commutation}
	\end{equation}
	\begin{equation}
	D_{u}\prod_{j=1}^{N}C\left(u_{j}\right)\left|0\right\rangle =d_{u}\prod_{j=1}^{N}\frac{1}{b_{ju}}C\left(u_{j}\right)\left|0\right\rangle +\sum_{j=1}^{N}d_{j}\frac{c_{uj}}{b_{uj}}C\left(u\right)\prod_{l=1\neq j}^{N}\frac{1}{b_{lj}}C\left(u_{l}\right)\left|\Downarrow\right\rangle ,\label{eq:D_string_commutation}
	\end{equation}
\end{widetext}where the vacuum eigenvalues of the operators, $A_{u}\left|\Downarrow\right\rangle =a_{u}\left|\Downarrow\right\rangle $
and $D_{u}\left|\Downarrow\right\rangle =d_{u}\left|\Downarrow\right\rangle $,
are 
\begin{equation}
a_{u}=\frac{\cosh^{L}\left(u-\eta\right)}{\cosh^{L}\left(u+\eta\right)}\quad\textrm{and}\quad d_{u}=1.\label{eq:ad_vac}
\end{equation}

Since the right hand side of Eqs. (\ref{eq:A_string_commutation},\ref{eq:D_string_commutation})
contains terms that are not proportional to the original state
multiplied by a scalar, an arbitrary Bethe state is not an eigenstate
of the transfer matrix $\tau$ for an arbitrary set of the auxiliary
parameters $u_{j}$. However, the second terms in Eqs. (\ref{eq:A_string_commutation},\ref{eq:D_string_commutation})
can be made zero by selecting specific sets of $u_{j}$ that are solutions
of the following set of non-linear equations,

\begin{equation}
\frac{a_{j}}{d_{j}}=\prod_{l=1\neq j}^{N}\frac{b_{jl}}{b_{lj}},
\end{equation}
where we have used the shorthand with the subscripts, i.e. $a_{j}\equiv a\left(u_{j}\right)$
and $b_{jl}\equiv b\left(u_{j}-u_{l}\right)$. Substitution of the
expressions for $a_{j}$ and $d_{j}$ from Eq. (\ref{eq:ad_vac})
and for $b_{jl}$ from Eq. (\ref{eq:R_matrix}) gives the following
Bethe equation and the eigenvalue of the transfer matrix $\tau$,
\begin{equation}
\frac{\cosh\left(u_{j}-\eta\right)^{L}}{\cosh\left(u_{j}+\eta\right)^{L}}=\prod_{l=1\neq j}^{N}\frac{\sinh\left(u_{j}-u_{l}-2\eta\right)}{\sinh\left(u_{j}-u_{l}+2\eta\right)},\label{eq:BA_ABA}
\end{equation}

\begin{equation}
\mathcal{T}_{u}=a_{u}\prod_{j=1}^{N}\frac{1}{b_{uj}}+d_{u}\prod_{j=1}^{N}\frac{1}{b_{ju}}.\label{eq:tau_eigenvalue}
\end{equation}
The Bethe ansatz equations \textendash{} in the coordinate
representation \textendash{} are obtained under substitution of 
\begin{equation}
u_{j}=\ln\left[\sqrt{\frac{1-e^{iq_{j}-2\eta}}{1-e^{-iq_{j}-2\eta}}}\right]-\frac{iq_{j}}{2}
\end{equation}
and 
\begin{equation}
\eta=\frac{\textrm{acosh}\Delta}{2}
\end{equation}
into Eq. (\ref{eq:BA_ABA}).

The scalar product between two Bethe states $\left\langle \mathbf{v}\right|$
and $\left|\mathbf{u}\right\rangle $ can be calculated using the
commutation relations in Eqs. (\ref{eq:ABA_BC}-\ref{eq:ABA_AD}).
The multiplication of the bra and ket states in the form of Eq. (\ref{eq:psi_ABA})
is evaluated by commuting each operator $B\left(v_{j}\right)$ from
left to right through the product of $C\left(u_{j}\right)$ operators
and then by using the vacuum eigenvalues of the generated $A$ and
$D$ operators from Eq. (\ref{eq:ad_vac}). When $u_{j}$ is a solution
of Eq. (\ref{eq:BA_ABA}) and $v_{j}$ is an arbitrary set of auxiliary
parameters the result can be written in a compact form as a determinant
of an $N\times N$ matrix \textendash{} the so-called Slavnov's formula, \cite{Slavnov89}

\begin{equation}
\left\langle \mathbf{v}|\mathbf{u}\right\rangle =\frac{\prod_{i,j=1}^{N}\sinh\left(v_{j}-u_{i}\right)}{\prod_{j<i}\sinh\left(v_{j}-v_{i}\right)\prod_{j<i}\sinh\left(u_{j}-u_{i}\right)}\det\hat{T},\label{eq:scalar_product}
\end{equation}
where matrix elements are $T_{ab}=\partial_{u_{a}}\mathcal{T}\left(v_{b}\right)$.
Under substitution of $\mathcal{T}\left(u\right)$ from Eq. (\ref{eq:R_matrix})
these matrix elements read \begin{widetext}
	\begin{equation}
	T_{ab}=\frac{\cosh^{L}\left(v_{b}-\eta\right)}{\cosh^{L}\left(v_{b}+\eta\right)}\frac{\sinh\left(2\eta\right)}{\sinh^{2}\left(v_{b}-u_{a}\right)}\prod_{j=1\neq a}^{N}\frac{\sinh\left(v_{b}-u_{j}+2\eta\right)}{\sinh\left(v_{b}-u_{j}\right)}-\frac{\sinh\left(2\eta\right)}{\sinh^{2}\left(u_{a}-v_{b}\right)}\prod_{j=1\neq a}^{N}\frac{\sinh\left(u_{j}-v_{b}+2\eta\right)}{\sinh\left(u_{j}-v_{b}\right)}.\label{eq:scalar_product_matrix_elements}
	\end{equation}
	
	The normalisation factor of Bethe states in Eq. (\ref{eq:psi_ABA})
	can be evaluated by taking the $\mathbf{v}\rightarrow\mathbf{u}$
	limit of Eq. (\ref{eq:scalar_product}), \cite{Korepin82_main,Gaudin81_main}
	\begin{equation}
	\left\langle \mathbf{u}|\mathbf{u}\right\rangle =\sinh^{N}\left(2\eta\right)\prod_{i\neq j=1}^{N}\frac{\sinh\left(u_{j}-u_{i}+2\eta\right)}{\sinh\left(u_{j}-u_{i}\right)}\det\hat{M},
	\end{equation}
	where the matrix elements are 
	\begin{equation}
	M_{ab}=\begin{cases}
	-L\frac{\sinh2\eta}{\cosh\left(u_{a}+\eta\right)\cosh\left(u_{a}-\eta\right)}-\sum_{j\neq a}\frac{\sinh4\eta}{\sinh\left(u_{a}-u_{j}-2\eta\right)\sinh\left(u_{a}-u_{j}+2\eta\right)} & ,a=b,\\
	\frac{\sinh4\eta}{\sinh\left(u_{b}-u_{a}+2\eta\right)\sinh\left(u_{b}-u_{a}-2\eta\right)} & ,a\neq b.
	\end{cases}
	\end{equation}
\end{widetext}

\section{Derivation of the dynamical matrix element for spins}
\setcounter{equation}{0}

In this section we will calculate the matrix element $\left\langle \mathbf{q}|\mathbf{S}_{1}\cdot\mathbf{S}_{2}|0\right\rangle $
\textendash{} with respect to Bethe states of the spin Hamiltonian
\textendash{} that is needed for evaluating Eq. (11).
We start by splitting the matrix element of the scalar product $\mathbf{S}_{1}\cdot\mathbf{S}_{2}$
into three parts, 
\begin{equation}
\left\langle \mathbf{q}|\mathbf{S}_{1}\cdot\mathbf{S}_{2}|0\right\rangle _{m}=G_{+-}+G_{-+}+G_{zz},\label{eq:matrix_elem_orig}
\end{equation}
where 
\begin{equation}
G_{+-}=\frac{1}{2}\left\langle \mathbf{v}\right|S_{1}^{+}S_{2}^{-}\left|\mathbf{u}\right\rangle ,\label{eq:Gpm_def}
\end{equation}
\begin{equation}
G_{-+}=\frac{1}{2}\left\langle \mathbf{v}\right|S_{1}^{-}S_{2}^{+}\left|\mathbf{u}\right\rangle ,\label{eq:Gmp_def}
\end{equation}
\begin{equation}
G_{zz}=\left\langle \mathbf{v}\right|S_{1}^{z}S_{2}^{z}\left|\mathbf{u}\right\rangle ,\label{eq:Gzz_def}
\end{equation}
$u_{j}$ are the quasimomenta of the ground state $\left|0\right\rangle $,
and $v_{j}$ are the quasimomenta of an excited state $\left|\mathbf{q}\right\rangle $
with the same number of particles.

The local spin operators of the model in Eq. (\ref{eq:Hm_app})  can be expressed in terms of the algebraic Bethe ansatz
operators from Eq. (\ref{eq:BA_operators}) as \cite{Kitanine00, Kitanine99, Maillet00}
\begin{eqnarray}
S_{1}^{+} & = & C_{\xi}\tau_{\xi}^{L-1},\;S_{2}^{+}=\tau_{\xi}C_{\xi}\tau_{\xi}^{L-2},\label{eq:Sp_ABA}\\
S_{1}^{-} & = & B_{\xi}\tau_{\xi}^{L-1},\;S_{2}^{-}=\tau_{\xi}B_{\xi}\tau_{\xi}^{L-2},\label{eq:Sm_ABA}\\
S_{1}^{z} & = & S_{2}^{z}\frac{A_{\xi}-D_{\xi}}{2}\tau{}_{\xi}^{L-1},\;S_{2}^{z}=\tau_{\xi}\frac{A_{\xi}-D_{\xi}}{2}\tau_{\xi}^{L-2},\hspace{7mm}\label{eq:Sz_ABA}
\end{eqnarray}
where $\xi=-i\pi/2+\eta$.

Firstly, we evaluate the $+-$ correlation function. Under the substitution
of Eqs.(\ref{eq:Sp_ABA}, \ref{eq:Sm_ABA}) in to Eq. (\ref{eq:Gpm_def})
it reads
\begin{equation}
G_{+-}=\frac{1}{2}\left\langle \mathbf{v}|C_{\xi}B_{\xi}|\mathbf{u}\right\rangle .
\end{equation}
Commuting of the $B_{\xi}$ operator from left to right through a
product of $C\left(u_{j}\right)$ operators by means of the commutation
relations in Eqs. (\ref{eq:ABA_BC}-\ref{eq:ABA_AD}) gives \begin{widetext}
	\begin{equation}
	B_{\xi}\prod_{j=1}^{N}C_{u_{j}}\left|\Downarrow\right\rangle =\sum_{x=1}^{N+1}a_{x}c_{x\xi}\prod_{i=1\neq x}^{N+1}\frac{1}{b_{xi}}\sum_{y=1\neq x}^{N+1}c_{\xi y}\prod_{j=1\neq x,y}^{N+1}\frac{1}{b_{jy}}\prod_{j=1\neq x,y}^{N+1}C_{u_{j}}\left|\Downarrow\right\rangle ,\label{eq:B_stringC_commutator}
	\end{equation}
	where $u_{N+1}\equiv\xi$. Multiplying the above expression by $C_{\xi}$
	and evaluating the scalar product with the final state $\left\langle \mathbf{v}\right|$
	we obtain
	
	\begin{eqnarray}
	G_{+-} & = & \frac{1}{2}\sum_{x=1}^{N}a_{x}\frac{c_{x\xi}}{b_{x\xi}}\prod_{i=1\neq x}^{N}\frac{1}{b_{xi}}\sum_{y=1\neq x}^{N}\frac{c_{\xi y}}{b_{\xi y}}\prod_{j=1\neq x,y}^{N}\frac{1}{b_{jy}}\left\langle u_{x-1},\xi,u_{x+1},u_{y-1},\xi,u_{y+1}|\mathbf{v}\right\rangle \nonumber \\
	&  & +\frac{1}{2}\sum_{x=1}^{N}a_{x}\frac{c_{x\xi}}{b_{x\xi}}\prod_{i=1\neq x}^{N}\frac{1}{b_{xi}}\prod_{j=1\neq x}^{N}\frac{1}{b_{j\xi}}\left\langle u_{x-1},\xi,u_{x+1}|\mathbf{v}\right\rangle .\label{eq:Gpm_result}
	\end{eqnarray}
\end{widetext}Here the property $\left\langle \mathbf{v}|\mathbf{u}\right\rangle =\left\langle \mathbf{u}|\mathbf{v}\right\rangle $
where $v_{j}$ satisfy the Bethe equations and $u_{j}$ is an arbitrary
set of auxiliary parameters \cite{Kitanine00,Kitanine99} was used. 

The remaining scalar product can be evaluated using the Slavnov's
formula  (\ref{eq:scalar_product}). By substituting $\xi=-i\pi/2+\eta$
into $\left\langle u_{x-1},\xi,u_{x+1}|\mathbf{v}\right\rangle $
in the second line of Eq. (\ref{eq:Gpm_result}) explicitly we obtain\begin{widetext}
	\begin{equation}
	\left\langle u_{x+1},\xi,u_{x-1}|\mathbf{v}\right\rangle =\frac{i\left(-1\right)^{x}\prod_{j}^{N}\cosh\left(v_{j}+\eta\right)\prod_{j,i\neq x}^{N}\sinh\left(u_{i}-v_{j}\right)\det\hat{T}^{\left(x\right)}}{\prod_{j\neq x}^{N}\cosh\left(u_{j}-\eta\right)\prod_{i<j}\sinh\left(v_{i}-v_{j}\right)\prod_{i<j\neq x}\sinh\left(u_{i}-u_{j}\right)},\label{eq:uv_xi}
	\end{equation}
\end{widetext}where the matrix elements are 
\begin{equation}
T_{ab}^{\left(x\right)}=T_{ab},\quad b\neq x,
\end{equation}
\begin{equation}
T_{ax}^{\left(x\right)}=\frac{\sinh\left(2\eta\right)}{\cosh\left(v_{a}-\eta\right)\cosh\left(v_{a}+\eta\right)},\quad b=x,
\end{equation}
and $T_{ab}$ are given in Eq. (\ref{eq:scalar_product_matrix_elements}).

Substitution of the two identical $u_{j}=u_{j'}=\xi$ into the scalar
product $\left\langle u_{x-1},\xi,u_{x+1},u_{y-1},\xi,u_{y+1}|\mathbf{v}\right\rangle $
in the first line in Eq. (\ref{eq:Gpm_result}) makes the prefactor
in Eq. (\ref{eq:scalar_product}) divergent, i.e. the prefactor has
a pole of the first order as a function of $\left(u_{j'}-u_{j}\right)$.
Simultaneously, the determinant in Eq. (\ref{eq:scalar_product}) becomes
zero under the same substitution $u_{j}=u_{j'}=\xi$ since two lines
of the matrix in Eq. (\ref{eq:scalar_product_matrix_elements}) becomes
identical. Thus, we will derive the explicit expression for the whole
scalar product by substituting $u_{j}=\xi$ first, then, by taking
the limit $u_{j'}=\bar{\xi}\rightarrow\xi$. Expanding the matrix
elements in Eq. (\ref{eq:scalar_product_matrix_elements}) in a Taylor
series in $\left(\bar{\xi}-\xi\right)$ and using general matrix identities
we obtain 
\begin{multline}
\left|\begin{array}{c}
\cdots\\
\mathbf{A}^{T}\\
\cdots\\
\mathbf{A}^{T}+\left(\beta\mathbf{A}^{T}+\mathbf{X}^{T}\right)\left(\bar{\xi}-\xi\right)\\
\cdots
\end{array}\right|=\left(\bar{\xi}-\xi\right)\left|\begin{array}{c}
\cdots\\
\mathbf{A}^{T}\\
\cdots\\
\mathbf{X}^{T}\\
\cdots
\end{array}\right|,\label{eq:jjp_matrix_identity}
\end{multline}
where 
\begin{equation}
A_{a}=\frac{\sinh\left(2\eta\right)}{\cosh\left(v_{a}-\eta\right)\cosh\left(v_{a}+\eta\right)}\prod_{j}^{N}\frac{\cosh\left(v_{j}+\eta\right)}{\cosh\left(v_{j}-\eta\right)}
\end{equation}
is the $j^{\textrm{th}}$ row of Eq. (\ref{eq:scalar_product_matrix_elements})
under the substitution $u_{j}=\xi$, 
\begin{equation}
X_{a}=\frac{\sinh2\eta\sinh2v_{a}}{\cosh^{2}\left(v_{a}-\eta\right)\cosh^{2}\left(v_{a}+\eta\right)}\prod_{j}^{N}\frac{\cosh\left(v_{j}+\eta\right)}{\cosh\left(v_{j}-\eta\right)},
\end{equation}
is the linear coefficient in the Taylor expansion of the $j'^{\th}$
row of Eq. (\ref{eq:scalar_product_matrix_elements}) around the point
$u_{j'}=\xi$, which is not collinear with $A_{a}$ in the vector
space, and $\beta$ is the part of the linear coefficient that is
collinear with $A_{a}$.

Cancellation of the $\bar{\xi}-\xi$ from the denominator in Eq.~(\ref{eq:jjp_matrix_identity})
with the $1/\left(\bar{\xi}-\xi\right)$ from the prefactor in Eq.~(\ref{eq:scalar_product}) makes the whole scalar product finite.
Contributions of the orders higher than one (in the expansion of the
determinant) vanish in the limit $\bar{\xi}\rightarrow\xi$ and we
obtain \begin{widetext}
	\begin{equation}
	\left\langle u_{x-1},\xi,u_{x+1},u_{y-1},\xi,u_{y+1}|\mathbf{v}\right\rangle =\left(-1\right)^{x+y}\frac{\prod_{j}\cosh^{2}\left(v_{j}+\eta\right)}{\prod_{j\neq x,y}\cosh^{2}\left(u_{j}-\eta\right)}\frac{\prod_{j,j';j'\neq x,y}\sinh\left(u_{j'}-v_{j}\right)\det\hat{T}^{\left(xy\right)}}{\prod_{j<j'}\sinh\left(v_{i}-v_{j}\right)\prod_{j<j'\neq x,y}\sinh\left(u_{i}-u_{j}\right)},\label{eq:uv_xixip}
	\end{equation}
\end{widetext}where the matrix elements are
\begin{equation}
T_{ab}^{\left(xy\right)}=\begin{cases}
T_{ab} & ,b\neq x,y,\\
T_{ab}^{\left(b\right)} & ,b=\min\left(x,y\right),\\
\frac{\sinh2\eta\sinh2v_{a}}{\cosh^{2}\left(v_{a}-\eta\right)\cosh^{2}\left(v_{a}+\eta\right)} & ,b=\max\left(x,y\right).
\end{cases}
\end{equation}

Secondly, we turn to evaluating the $-+$ correlation function. Under
the substitution of Eqs.(\ref{eq:Sp_ABA}, \ref{eq:Sm_ABA}) into
Eq. (\ref{eq:Gmp_def}) it reads
\begin{equation}
G_{-+}=\frac{1}{2}\left\langle \mathbf{v}|B_{\xi}C_{\xi}|\mathbf{u}\right\rangle .\label{eq:Gmp}
\end{equation}
When $B_{\xi}$ is commuted through the product of $C_{u_{j}}$ operators
using the general result in Eq. (\ref{eq:B_stringC_commutator}),
the first step of commuting $B_{\xi}$ with $C_{\xi}$ introduces
a divergent denominator through the commutation relation in Eq. (\ref{eq:ABA_BC}).
However, the operator factor in the numerator of Eq.~(\ref{eq:ABA_BC})
becomes zero at the same time making the whole expression finite.
Since the divergence occurs at the first step of commuting $B_{\xi}$
through a product of $N+1$ operators $C\left(u_{j}\right)$, taking
the limit after using Eq. (\ref{eq:B_stringC_commutator}), as it
is done in Ref. \onlinecite{Kitanine00}, creates an extra and significant
computation problems: the original divergence spreads through many
terms under the  sum in Eq. (\ref{eq:B_stringC_commutator}) and
cancelling them explicitly is a complicated problem. 

Here we will do it in a different way by cancelling this intermediate
divergence from the beginning in Eq. (\ref{eq:Gmp}). Expanding the
numerator and the denominator of the commutation relation in Eq. (\ref{eq:ABA_BC})
up to the linear order in $\bar{\xi}-\xi$, where $u\rightarrow\overline{\xi}$
and $v\rightarrow\xi$ auxiliary parameters were relabeled, we cancel
the $\bar{\xi}-\xi$ with $1/(\bar{\xi}-\xi)$. Then, substituting the
result of this procedure in Eq. (\ref{eq:Gmp}) we obtain 
\begin{multline}
G_{-+}=\frac{1}{2}\left\langle \mathbf{v}|C_{\xi}B_{\xi}|\mathbf{u}\right\rangle +\frac{\sinh2\eta}{2}\\
\times\lim_{\bar{\xi}\rightarrow\xi}\partial_{\bar{\xi}}\left(\left\langle \mathbf{v}|A_{\bar{\xi}}D_{\xi}|\mathbf{u}\right\rangle -\left\langle \mathbf{v}|A_{\xi}D_{\bar{\xi}}|\mathbf{u}\right\rangle \right),\label{eq:Gmp_diff}
\end{multline}
where $\left\langle \mathbf{v}|C_{\xi}B_{\xi}|\mathbf{u}\right\rangle $
has already been calculated in Eq.~(\ref{eq:Gpm_result}). 

The remaining two correlation functions under the derivative in Eq.
(\ref{eq:Gmp_diff}) can be calculated by successive use of the general
result of commuting $A_{u}$ and $D_{v}$ operators through a product
of $C\left(u_{j}\right)$ operators in Eqs. (\ref{eq:A_string_commutation},
\ref{eq:D_string_commutation}). The scalar product of $\left\langle \mathbf{v}\right|$
with the result of the commutation procedure gives\begin{widetext}
	\begin{eqnarray}
	\left\langle \mathbf{v}|A_{\bar{\xi}}D_{\xi}|\mathbf{u}\right\rangle  & = & a_{\bar{\xi}}\prod_{l=1}^{N}\frac{1}{b_{l\xi}}\prod_{j=1}^{N}\frac{1}{b_{\bar{\xi}j}}\delta_{\mathbf{u},\mathbf{v}}-\prod_{l=1}^{N}\frac{1}{b_{l\xi}}\sum_{j=1}^{N}a_{j}\frac{c_{\bar{\xi}j}}{b_{\bar{\xi}j}}\prod_{l=1\neq j}^{N}\frac{1}{b_{jl}}\left\langle u_{j-1},\bar{\xi},u_{j+1}|\mathbf{v}\right\rangle \nonumber \\
	&  & +\sum_{j=1}^{N}\frac{c_{\xi j}}{b_{\xi j}}\prod_{l=1\neq j}^{N}\frac{1}{b_{lj}}a_{\bar{\xi}}\frac{1}{b_{\bar{\xi}\xi}}\prod_{l=1\neq j}^{N}\frac{1}{b_{\bar{\xi}l}}\left\langle u_{j-1},\xi,u_{j+1}|\mathbf{v}\right\rangle \nonumber \\
	&  & -\sum_{j=1}^{N}\frac{c_{\xi j}}{b_{\xi j}}\prod_{l=1\neq j}^{N}\frac{1}{b_{lj}}\sum_{j'=1\neq j}^{N}a_{j'}\frac{c_{\bar{\xi}j'}}{b_{\bar{\xi}j'}}\frac{1}{b_{j'\xi}}\prod_{l=1\neq j,j'}^{N}\frac{1}{b_{j'l}}\left\langle u_{j-1},\xi,u_{j+1},u_{j'-1},\bar{\xi}u_{j'+1}|\mathbf{v}\right\rangle \nonumber \\
	&  & -\sum_{j=1}^{N}\frac{c_{\xi j}}{b_{\xi j}}\prod_{l=1\neq j}^{N}\frac{1}{b_{lj}}a_{\xi}\frac{c_{\bar{\xi}\xi}}{b_{\bar{\xi}\xi}}\prod_{l=1\neq j}^{N}\frac{1}{b_{\xi l}}\left\langle u_{j-1},\bar{\xi},u_{j+1}|\mathbf{v}\right\rangle ,\label{eq:AD_xipxi}
	\end{eqnarray}
	\begin{eqnarray}
	\left\langle \mathbf{v}|A_{\xi}D_{\bar{\xi}}|\mathbf{u}\right\rangle  & = & a_{\bar{\xi}}\prod_{l=1}^{N}\frac{1}{b_{l\xi}}\prod_{j=1}^{N}\frac{1}{b_{\bar{\xi}j}}\delta_{\mathbf{u},\mathbf{v}}-\prod_{l=1}^{N}\frac{1}{b_{l\xi}}\sum_{j=1}^{N}a_{j}\frac{c_{\bar{\xi}j}}{b_{\bar{\xi}j}}\prod_{l=1\neq j}^{N}\frac{1}{b_{jl}}\left\langle u_{j-1},\bar{\xi},u_{j+1}|\mathbf{v}\right\rangle \nonumber \\
	&  & +\sum_{j=1}^{N}\frac{c_{\xi j}}{b_{\xi j}}\prod_{l=1\neq j}^{N}\frac{1}{b_{lj}}a_{\bar{\xi}}\frac{1}{b_{\bar{\xi}\xi}}\prod_{l=1\neq j}^{N}\frac{1}{b_{\bar{\xi}l}}\left\langle u_{j-1},\xi,u_{j+1}|\mathbf{v}\right\rangle \nonumber \\
	&  & -\sum_{j=1}^{N}\frac{c_{\xi j}}{b_{\xi j}}\prod_{l=1\neq j}^{N}\frac{1}{b_{lj}}\sum_{j'=1\neq j}^{N}a_{j'}\frac{c_{\bar{\xi}j'}}{b_{\bar{\xi}j'}}\frac{1}{b_{j'\xi}}\prod_{l=1\neq j,j'}^{N}\frac{1}{b_{j'l}}\left\langle u_{j-1},\xi,u_{j+1},u_{j'-1},\bar{\xi}u_{j'+1}|\mathbf{v}\right\rangle \label{eq:AD_xixip}
	\end{eqnarray}
	for the both terms in the second line of Eq. (\ref{eq:Gmp_diff})
	respectively. Then, after taking the derivative of Eqs. (\ref{eq:AD_xipxi},
	\ref{eq:AD_xixip}), with respect to $\bar{\xi}$ and the limit $\bar{\xi}\rightarrow\xi$,
	we substitute the results in to Eq. (\ref{eq:Gmp_diff}) and obtain
	\begin{align}
	G_{-+} & =G_{+-}+\frac{\sinh2\eta}{2}\Bigg[\prod_{l=1}^{N}\frac{1}{b_{l\xi}}\sum_{j=1}^{N}\sum_{l=1}^{N}\left[\tanh\left(u_{l}+\eta\right)-\tanh\left(v_{l}-\eta\right)\right]a_{j}\frac{c_{j\xi}}{b_{j\xi}}\prod_{l=1\neq j}^{N}\frac{1}{b_{jl}}\left\langle u_{j-1},\xi,u_{j+1}|\mathbf{v}\right\rangle \nonumber \\
	& +\sum_{j=1}^{N}\frac{c_{j\xi}}{b_{j\xi}}\prod_{l=1\neq j}^{N}\frac{1}{b_{lj}}\sum_{j'=1\neq j}^{N}a_{j'}\frac{c_{\xi j'}}{b_{\xi j'}}\left(\tanh\left(u_{j'}+\eta\right)-\tanh\left(u_{j}-\eta\right)\right)\frac{1}{b_{j'\xi}}\prod_{l=1\neq j,j'}^{N}\frac{1}{b_{j'l}}\left\langle u_{j-1},\xi,u_{j+1},u_{j'-1},\xi u_{j'+1}|\mathbf{v}\right\rangle \nonumber \\
	& +\prod_{l=1}^{N}\frac{1}{b_{l\xi}}\sum_{j=1}^{N}a_{j}\frac{c_{j\xi}}{b_{j\xi}}\prod_{l=1\neq j}^{N}\frac{1}{b_{jl}}\left\langle u_{j-1},\xi,u_{j+1}|\mathbf{v}\right\rangle '\Bigg],\label{eq:Gmp_result}
	\end{align}
	where the derivative of $\left\langle u_{j-1},\bar{\xi},u_{j+1}|\mathbf{v}\right\rangle $
	with respect to $\bar{\mathbf{\xi}}$ results in an additional determinant,
	\begin{equation}
	\left\langle u_{x-1},\xi,u_{x+1}|\mathbf{v}\right\rangle '=\frac{i\left(-1\right)^{j}\prod_{j'}^{N}\cosh\left(v_{j'}+\eta\right)\prod_{j',i\neq x}^{N}\sinh\left(u_{i}-v_{j'}\right)\det\hat{T}^{'\left(x\right)}}{\prod_{j'\neq x}^{N}\cosh\left(u_{j'}-\eta\right)\prod_{i<j'}\sinh\left(v_{i}-v_{j'}\right)\prod_{i<j'\neq x}\sinh\left(u_{i}-u_{j'}\right)},
	\end{equation}
	where the matrix elements are 
	\begin{equation}
	T_{ax}^{'\left(x\right)}=\frac{2\sinh2\eta\tanh\left(v_{a}-\eta\right)}{\cosh\left(v_{a}-\eta\right)\cosh\left(v_{a}+\eta\right)}-\frac{\sinh^{2}2\eta}{\cosh\left(v_{a}-\eta\right)\cosh\left(v_{a}+\eta\right)}\sum_{j=1\neq a}^{N}\frac{1}{\cosh\left(v_{j}-\eta\right)\cosh\left(v_{j}+\eta\right)}
	\end{equation}
	for $b=x$ and the remaining entries for $b\neq x$ are $T_{ab}^{'\left(x\right)}\equiv T_{ab}$
	from Eq. (\ref{eq:scalar_product_matrix_elements}).\end{widetext}

Thirdly, we evaluate the $zz$ correlation function. Under the substitution
of Eq. (\ref{eq:Sz_ABA}) in Eq. (\ref{eq:Gzz_def}) it reads
\begin{equation}
G_{zz}=\frac{1}{4}\left\langle \mathbf{v}\left|\left(A_{\xi}-D_{\xi}\right)\left(A_{\xi}-D_{\xi}\right)\right|\mathbf{u}\right\rangle .
\end{equation}
Before proceeding with the commutation procedure as in the two previous
cases we rewrite the above expression in a form more convenient for
such a calculation using the definition of the transfer matrix, $A_{\xi}-D_{\xi}=2A_{\xi}-\tau_{\xi}$,
and its eigenvalue $\tau_{\xi}\left|\mathbf{u}\right\rangle =\mathcal{T}_{\xi}\left|\mathbf{u}\right\rangle $,
\begin{equation}
G_{zz}=\frac{1}{2}\left\langle \mathbf{v}\left|A_{\xi}^{2}-\mathcal{T}_{\xi}A_{\xi}-D_{\xi}A_{\xi}\right|\mathbf{u}\right\rangle ,\label{eq:Gzz_tranformed}
\end{equation}
where $\mathcal{T}_{\xi}=\prod_{j}b_{j\xi}^{-1}$ is given by Eq.
(\ref{eq:tau_eigenvalue}) and we have assumed that $\left\langle \mathbf{v}\right|$
and $\left|\mathbf{u}\right\rangle $ are a pair of orthogonal eigenstates,
i.e. $\left\langle \mathbf{v}|\mathbf{u}\right\rangle =0$.

The correlation function of $A_{\xi}$ and $A_{\xi}^{2}$ can be calculated
using Eq. (\ref{eq:A_string_commutation}) once and twice respectively.
The scalar products of $\left\langle \mathbf{v}\right|$ with the
results of the commutation procedures are 
\begin{equation}
\left\langle \mathbf{v}\left|A_{\xi}\right|\mathbf{u}\right\rangle =-\sum_{x=1}^{N}a_{j}\frac{c_{\xi x}}{b_{\xi x}}\prod_{l=1\neq j}^{N}\frac{1}{b_{xl}}\left\langle u_{x+1},\xi,u_{x-1}|\mathbf{v}\right\rangle ,\label{eq:A_corr}
\end{equation}
\begin{widetext}
	\begin{equation}
	\left\langle \mathbf{v}\left|A_{\xi}^{2}\right|\mathbf{u}\right\rangle =4\sum_{x=1}^{N}a_{x}\frac{c_{\xi x}}{b_{\xi x}}\prod_{l;l\neq x}^{N}\frac{1}{b_{xl}}\sum_{y;y\neq x}^{N}a_{y}\frac{c_{\xi y}}{b_{\xi y}}\prod_{l;l\neq x,y}^{N}\frac{1}{b_{xl}}\frac{1}{b_{y\xi}}\left\langle u_{x-1},\xi,u_{x+1},u_{y-1},\xi,u_{y+1}|\mathbf{v}\right\rangle ,\label{eq:A2_corr}
	\end{equation}
	where the scalar products in the right hand sides are already given
	in Eqs. (\ref{eq:uv_xi}, \ref{eq:uv_xixip}) in explicit form.
	
	Evaluation of the remaining $D_{\xi}A_{\xi}$ correlation function
	involves the same problem of taking the limit $v\rightarrow u=\xi$
	in commutation relation Eq. (\ref{eq:ABA_AD}), as in the calculation
	of the $-+$ correlation function. Here we resolve it in the same
	way as we have already done in evaluating Eq. (\ref{eq:Gmp}). Expanding
	the numerator and the denominator of Eq. (\ref{eq:ABA_AD}) in $v-u\rightarrow0$
	we rewrite the $D_{\xi}A_{\xi}$ correlation function as 
	\begin{equation}
	\left\langle \mathbf{v}|D_{\xi}A_{\xi}|\mathbf{u}\right\rangle =\left\langle \mathbf{v}|\mathcal{T}_{\xi}A_{\xi}-A_{\xi}^{2}|\mathbf{u}\right\rangle -\sinh2\eta\lim_{\bar{\xi}\rightarrow\xi}\partial_{\bar{\xi}}\left\langle \mathbf{v}|C_{\xi}B_{\bar{\xi}}-C_{\bar{\xi}}B_{\xi}|\mathbf{u}\right\rangle .\label{eq:DA_diff}
	\end{equation}
	We use the general result in Eq. (\ref{eq:B_stringC_commutator})
	and write the $C_{\xi}B_{\bar{\xi}}$ and $C_{\bar{\xi}}B_{\xi}$
	correlation functions under the derivative as 
	\begin{eqnarray}
	\left\langle \mathbf{v}|C_{\bar{\xi}}B_{\xi}|\mathbf{u}\right\rangle  & = & \sum_{x=1}^{N}a_{x}\frac{c_{x\xi}}{b_{x\xi}}\prod_{i=1\neq x}^{N}\frac{1}{b_{xi}}\sum_{y=1\neq x}^{N}d_{y}\frac{c_{\xi y}}{b_{\xi y}}\prod_{j=1\neq x,y}^{N}\frac{1}{b_{jy}}\left\langle u_{x-1},\bar{\xi},u_{x+1},u_{y-1},\xi,u_{y+1}|\mathbf{v}\right\rangle \nonumber \\
	&  & +\sum_{x=1}^{N}a_{x}\frac{c_{x\xi}}{b_{x\xi}}\prod_{i=1\neq x}^{N}\frac{1}{b_{xi}}\prod_{j=1\neq x}^{N}\frac{1}{b_{j\xi}}\left\langle u_{x-1},\bar{\xi},u_{x+1}|\mathbf{v}\right\rangle,
	\end{eqnarray}
	\begin{eqnarray}
	\left\langle \mathbf{v}|C_{\xi}B_{\bar{\xi}}|\mathbf{u}\right\rangle  & = & \sum_{x=1}^{N}a_{x}\frac{c_{x\bar{\xi}}}{b_{x\bar{\xi}}}\prod_{i=1\neq x}^{N}\frac{1}{b_{xi}}\sum_{y=1\neq x}^{N}d_{y}\frac{c_{\bar{\xi}y}}{b_{\bar{\xi}y}}\prod_{j=1\neq x,y}^{N}\frac{1}{b_{jy}}\left\langle u_{x-1},\xi,u_{x+1},u_{y-1},\bar{\xi},u_{y+1}|\mathbf{v}\right\rangle \nonumber \\
	&  & +\sum_{x=1}^{N}a_{x}\frac{c_{x\bar{\xi}}}{b_{x\bar{\xi}}}\prod_{i=1\neq x}^{N}\frac{1}{b_{xi}}\prod_{j=1\neq x}^{N}\frac{1}{b_{j\bar{\xi}}}\left\langle u_{x-1},\xi,u_{x+1}|\mathbf{v}\right\rangle .
	\end{eqnarray}
	
	Then, taking the derivative over $\bar{\xi}$, the limit $\bar{\xi}\rightarrow\xi$,
	and substituting the pair of the expressions above in Eqs. (\ref{eq:Gzz_tranformed},
	\ref{eq:DA_diff}), together with the expressions in Eqs. (\ref{eq:A_corr},
	\ref{eq:A2_corr}), we obtain
	\begin{align}
	G_{zz} & =\prod_{j}^{N}\frac{1}{b_{j\xi}}\sum_{j=1}^{N}a_{j}\frac{c_{\xi j}}{b_{\xi j}}\prod_{l=1\neq j}^{N}\frac{1}{b_{jl}}\left\langle u_{j+1},\xi,u_{j-1}|\mathbf{v}\right\rangle +\sum_{j=1}^{N}a_{j}\frac{c_{\xi j}}{b_{\xi j}}\prod_{l=1\neq j}^{N}\frac{1}{b_{jl}}\sum_{j'=1\neq j}^{N}a_{j'}\frac{c_{\xi j'}}{b_{\xi j'}}\prod_{l=1\neq j,j'}^{N}\frac{1}{b_{j'l}}\nonumber \\
	& \times\frac{1}{b_{j'\xi}}\left\langle u_{j-1},\xi,u_{j+1},u_{j'-1},\xi,u_{j'+1}|\mathbf{v}\right\rangle +\frac{\sinh2\eta}{2}\Bigg[\sum_{x=1}^{N}a_{x}\frac{c_{x\xi}}{b_{x\xi}}\prod_{i=1\neq x}^{N}\frac{1}{b_{xi}}\sum_{y=1\neq x}^{N}\frac{c_{\xi y}}{b_{\xi y}}\nonumber \\
	& \times\left(\tanh\left(u_{x}-\eta\right)+\tanh\left(u_{y}-\eta\right)\right)\prod_{j=1\neq x,y}^{N}\frac{1}{b_{jy}}\left\langle u_{x-1},\xi,u_{x+1},u_{y-1},\xi,u_{y+1}|\mathbf{v}\right\rangle \nonumber \\
	& +\sum_{x=1}^{N}a_{x}\frac{c_{x\xi}}{b_{x\xi}}\prod_{i=1\neq x}^{N}\frac{1}{b_{xi}}\prod_{j=1\neq x}^{N}\frac{1}{b_{j\xi}}\Big[\tanh\left(u_{x}-\eta\right)+\tanh\left(v_{x}-\eta\right)\nonumber \\
	& +\sum_{j'=1\neq x}^{N}\left[\tanh\left(v_{j'}-\eta\right)-\tanh\left(u_{j'}+\eta\right)\right]\Bigg]\left\langle u_{x-1},\xi,u_{x+1}|\mathbf{v}\right\rangle \nonumber \\
	& -\sum_{x=1}^{N}a_{x}\frac{c_{x\xi}}{b_{x\xi}}\prod_{i=1\neq x}^{N}\frac{1}{b_{xi}}\prod_{j=1\neq x}\frac{1}{b_{j\xi}}\left\langle u_{x-1},\xi,u_{x+1}|\mathbf{v}\right\rangle '\Bigg],\label{eq:Gzz_result}
	\end{align}
	where all scalar products are already given in explicit form above.
	
	Finally, we substitute Eqs. (\ref{eq:Gpm_result}, \ref{eq:Gmp_result},
	\ref{eq:Gzz_result}) in Eq. (\ref{eq:matrix_elem_orig}), rewrite
	the result in a more compact form by collecting similar terms, and
	use a general matrix identity $\det\hat{T}+\sum_{x=1}^{N}\det\hat{T}^{\left(x\right)}=\det\left(\hat{T}+\hat{X}\right)$,
	where the matrix $T^{\left(b\right)}$ is obtained by substitution
	of the $x^{th}$ column from the matrix $\hat{X}$ and rank of $\hat{X}$
	is equal to one. After constructing the corresponding matrices $\hat{X}$
	for a single sum over $x$ and for a sum over only one variable in the double
	sum over $x,y$ we obtain
	\begin{multline}
	\left\langle \mathbf{q}|\mathbf{S}_{1}\cdot\mathbf{S}_{2}|0\right\rangle =\frac{\prod_{j}^{N}\cosh\left(v_{j}+\eta\right)}{\prod_{i<j}\sinh\left(v_{i}-v_{j}\right)}\sum_{x=1}^{N}\left(-1\right)^{x}\frac{\prod_{i,j;j\neq x}^{N}\sinh\left(u_{j}-v_{i}\right)}{\prod_{j}^{N}\cosh^{2}\left(u_{j}-\eta\right)}\prod_{l=1\neq x}^{N}\frac{\sinh\left(u_{l}-u_{x}+2\eta\right)}{\sinh\left(u_{l}-u_{x}\right)}\Bigg[\det\hat{K}^{\left(x\right)}\\
	-\left(1-\frac{\sinh(2\eta)\sinh\eta\sinh u_{x}\prod_{j;j\neq x}^{N}\cosh\left(u_{j}+\eta\right)}{\prod_{i<j\neq x}\sinh\left(u_{i}-u_{j}\right)}\right)\det\hat{G}^{\left(x\right)}\Bigg]-\frac{\prod_{j}\cosh\left(u_{j}+\eta\right)\prod_{j}\cosh\left(v_{j}+\eta\right)}{\prod_{j}\cosh^{2}\left(u_{j}-\eta\right)\prod_{i<j}\sinh\left(v_{i}-v_{j}\right)}\det\hat{K},\label{eq:matrix_elem_result}
	\end{multline}
	where the matrix elements are
	\begin{align}
	K_{ab} & =T_{ab}+\left(-1\right)^{b}\frac{\sinh^{3}(2\eta)\sinh\eta\sinh u_{b}}{\cosh\left(u_{b}+\eta\right)}\frac{\prod_{j,i\neq b}^{N}\sinh\left(u_{i}-v_{j}\right)}{\prod_{i<j\neq b}\sinh\left(u_{i}-u_{j}\right)}\prod_{l=1\neq b}^{N}\frac{\sinh\left(u_{l}-u_{b}+2\eta\right)}{\sinh\left(u_{l}-u_{b}\right)}\nonumber \\
	& \times\frac{\frac{\sinh u_{b}}{\cosh\left(u_{b}+\eta\right)\cosh\eta}+\sum_{l=1}^{N}\left[\tanh\left(v_{l}+\eta\right)-\tanh\left(u_{l}+\eta\right)\right]}{\cosh\left(v_{a}-\eta\right)\cosh\left(v_{a}+\eta\right)},
	\end{align}
	\begin{equation}
	T_{ab}=\frac{\cosh^{L}\left(v_{b}-\eta\right)}{\cosh^{L}\left(v_{b}+\eta\right)}\frac{\sinh\left(2\eta\right)}{\sinh^{2}\left(v_{b}-u_{a}\right)}\prod_{j=1\neq a}^{N}\frac{\sinh\left(v_{b}-u_{j}+2\eta\right)}{\sinh\left(v_{b}-u_{j}\right)}-\frac{\sinh\left(2\eta\right)}{\sinh^{2}\left(u_{a}-v_{b}\right)}\prod_{j=1\neq a}^{N}\frac{\sinh\left(u_{j}-v_{b}+2\eta\right)}{\sinh\left(u_{j}-v_{b}\right)},
	\end{equation}
	\begin{multline}
	K_{ab}^{(x)}=T_{ab}+\frac{\left(-1\right)^{b}\sinh^{3}\left(2\eta\right)\textrm{sgn}\left(x-b\right)}{\cosh\left(v_{a}-\eta\right)\cosh\left(v_{a}+\eta\right)}\prod_{l=1\neq x,b}^{N}\frac{\sinh\left(u_{l}-u_{b}+2\eta\right)}{\sinh\left(u_{l}-u_{b}\right)}\frac{\cosh\left(u_{b}+\eta\right)\cosh\left(u_{x}-\eta\right)}{\prod_{i}^{N}\sinh\left(u_{b}-v_{i}\right)\prod_{i<j\neq x,b}\sinh\left(u_{i}-u_{j}\right)}\\
	\left(\frac{\cosh\left(u_{b}-\eta\right)}{\cosh\left(u_{b}+\eta\right)}-\frac{\sinh\left(u_{x}-u_{b}+2\eta\right)}{\sinh\left(u_{x}-u_{b}-2\eta\right)}+\frac{\sinh2\eta\cosh\left(u_{b}-2\eta\right)\sinh u_{x}}{\cosh\left(u_{x}-\eta\right)\cosh\left(u_{b}+\eta\right)}\right),\label{eq:Kxab_app}
	\end{multline}
	when $b\neq x$,
	\begin{equation} K_{ax}^{\left(x\right)}=\frac{\sinh(2\eta)\allowbreak\sinh(2v_{a})}{\cosh^{2}\left(v_{a}-\eta\right)\allowbreak\cosh^{2}\left(v_{a}+\eta\right)}
	\end{equation}
\end{widetext}
when $b=x$, $G_{ab}^{\left(x\right)}=T_{ab}$ when $b\neq x$, 
and
$G_{ax}^{\left(x\right)}=K_{ax}^{\left(x\right)}$ when $b=x$. The
result in Eq. (\ref{eq:matrix_elem_result}) was checked numerically
for $N=2,3$ using the direct summation over the spacial coordinates.
Eqs. (\ref{eq:matrix_elem_result}-\ref{eq:Kxab_app}) are Eqs. (\ref{eq:S1S2_dynamic_matrix_element}-\ref{eq:Kyay}).


\begin{thebibliography}{10}
\bibitem{anticrossing}The width of the anti-crossing is estimated
as a matrix element of the magnetostrictive interaction in Eq. (\ref{eq:V}),
$J_{1}\sqrt{\hbar b/\left(m\omega_{D}\right)}/L^{4}$ (where the spin matrix element belongs the second level of the hierarchy
$\sim1/L^{4}$) times the superradiant enhancement factor $\left(\omega_{D}L/J\right)^{3/2}$
since the number of the almost degenerate levels is $\sim\left(\omega_{D}L/J\right)^{3}$.
\bibitem{parity_note} In this paper we assume that $N$ is even obtaining $q_j=2\pi(I_j+1/2)/L$ in the $\Delta\rightarrow 0$ limit. For odd $N$s the $\Delta\rightarrow 0$ limit of Eqs. (A2, A3) gives $q_j=2\pi I_j/L$ that is a parity effect. We do not consider it here.
\end{thebibliography}
\end{document}